\documentclass[12pt]{article}
\pdfoutput=1
\usepackage{jcappub}

\usepackage{amsmath,amsfonts,amssymb,mathrsfs,graphicx,color,longtable,bm}
\usepackage{hyperref}
\usepackage{graphicx}
\usepackage{array}
\usepackage{enumitem}
\usepackage[explicit]{titlesec}
\usepackage[utf8]{inputenc}
\usepackage[normalem]{ulem}
\usepackage{xcolor}

\hypersetup{
     colorlinks   = true,
     citecolor    = red,
     urlcolor     = blue,
     linkcolor    = blue
}

\newcommand{\dd}{\mathrm{d}}

\makeatletter
\gdef\@fpheader{}
\makeatother

\begin{document}

\title{Chasing cosmic inflation: constraints for inflationary models and reheating insights}

\author[a,b,c]{Mario Ballardini}

\affiliation[a]{Dipartimento di Fisica e Scienze della Terra, Universit\`a degli Studi di Ferrara, via Giuseppe Saragat 1, 44122 Ferrara, Italy}
\affiliation[b]{INFN, Sezione di Ferrara, via Giuseppe Saragat 1, 44122 Ferrara, Italy}
\affiliation[c]{INAF/OAS Bologna, via Piero Gobetti 101, 40129 Bologna, Italy}

\emailAdd{mario.ballardini@unife.it}

\abstract{We investigate the impact of different choice of prior's range for the reheating epoch on cosmic inflation parameter inference 
in light of cosmic microwave background (CMB) anisotropy measurements from the {\em Planck} 2018 legacy release 
in combination with BICEP/Keck Array 2018 data and additional late-time cosmological observations such as uncalibrated 
Type Ia supernovae from the Pantheon catalogue, baryon acoustic oscillations and redshift space distortions 
from SDSS/BOSS/eBOSS. 
Here, we explore in particular the implications for the combination of reheating and inflationary-model parameter space 
considering $R+R^2$ inflation and a broad class of $\alpha$-attractor and D-brane models. 
Propagating the uncertainties due to an unknown reheating phase, these inflationary models completely cover the $n_{\rm s}$-$r$ parameter space allowed by {\em Planck} and BICEP/Keck data 
and represent good targets for future CMB and large-scale structure experiments.
We perform a Bayesian model comparison of inflationary models, taking into account the reheating uncertainties 
assuming a conservative but accurate modelling of inflationary predictions. 
$R+R^2$ inflation, T-model $\alpha$-attractor inflation for $n=1$, E-model $\alpha$-attractor inflation for $n=1/2$, 
and KKLT inflation for $p=5$ are the better performing models, with none being preferred at a statistically significant level.}

\maketitle

\section{Introduction}
Cosmic inflation \cite{Starobinsky:1980te,Guth:1980zm,Linde:1981mu,Albrecht:1982wi,Hawking:1982ga,Linde:1983gd} 
postulates an epoch of accelerated expansion in the very early Universe that flattens the spatial geometry and dilutes 
troublesome pre-inflationary relics. In its simplest realisation, the nearly exponential expansion is driven by a scalar field 
$\phi$, known as {\em inflaton}, slowly rolling down a sufficiently flat potential $V(\phi)$. During inflation, quantum fluctuations in the scalar field 
and in the metric are amplified and stretched to density fluctuations and gravitational waves on cosmological scales, respectively.

Measurements of cosmic microwave background (CMB) anisotropies, such as those from the \textit{Planck} satellite 
\cite{Planck:2013jfk,Planck:2013wtn,Planck:2015sxf,Planck:2015zfm,Planck:2018jri,Planck:2019kim}, have significantly contributed to observational 
constraints on cosmic inflation. The tight CMB constraints on spatial curvature, isocurvature fluctuations, and primordial non-Gaussianity all 
agree with expectations from the canonical single-field slow-roll (SFSR) inflationary paradigm.

Inflation also predicts a B-mode pattern in the CMB polarisation 
\cite{Kamionkowski:1996ks,Kamionkowski:1996zd,Seljak:1996gy,Zaldarriaga:1996xe,Seljak:1996ti} from inflationary primordial 
gravitational waves \cite{Starobinsky:1979ty,Rubakov:1982df,Fabbri:1983us,Abbott:1984fp} that will be hotly pursued by the 
next-generation CMB experiments \cite{Kamionkowski:1999qc,Kamionkowski:2015yta,SimonsObservatory:2018koc,CMB-S4:2020lpa,LiteBIRD:2022cnt}.

The combination of constraints on the scalar spectral index $n_{\rm s}$ and the tensor-to-scalar ratio $r$ can be used to 
discriminate between different inflation models. The predictions of any model depend on the number of $e$-folds 
$N_k \equiv \ln(a_{\rm end}/a_k)$ between the moment of horizon crossing for a mode with comoving wavenumber $k$, determined 
by $k=aH$, and the end of inflation. Determining the appropriate number of $e$-folds is necessary to accurately connect the features of 
the inflationary potential with cosmological observations 
\cite{Dodelson:2003vq,Liddle:2003as,Kinney:2005in,Martin:2006rs,Adshead:2010mc,Mortonson:2010er}.

The duration of the reheating phase is another essential ingredient in comparing theory with measurements. At the end of 
inflation, the inflaton field loses its energy, eventually initiating the radiation-dominated phase. In the simplest picture, 
this process is assumed to be instantaneous. In general, the physics of reheating is expected to be more complicated (see, 
e.g., Ref.~\cite{Allahverdi:2010xz} for a review), and it is usually described phenomenologically by two parameters: the 
number $N_{\rm re} \equiv \ln(a_{\rm re}/a_{\rm end})$ of $e$-folds between the end of inflation and the beginning of the 
radiation phase, and $\bar{w}_{\rm re}$, the average equation-of-state parameter during the reheating phase. Our imprecise 
knowledge of the physics of reheating introduces uncertainty into the derivation of constraints on the inflationary potential 
from measurements of $n_{\rm s}$ and $r$, dubbed as {\em reheating uncertainties} \cite{Kinney:2005in}. However, there are prospects for using 
observations to constrain reheating, particularly to constrain the reheating temperature, when all the energy of the inflaton 
field is converted to radiation 
\cite{Martin:2010kz,Martin:2010hh,Easther:2011yq,Planck:2013jfk,Martin:2013nzq,Martin:2014nya,Planck:2015sxf}. Constraining 
the reheating temperature enables the testing and selection of inflation models with future CMB experiments 
\cite{Martin:2014rqa}.

More precisely, constraints on reheating are derived by requiring that the number of $e$-folds between the time that the 
current comoving horizon scale exited the horizon during inflation and the end of inflation must be related to the number of 
$e$-folds between the end of inflation and today. By imposing this requirement, several groups have derived constraints on the 
reheating parameter space and the inflationary-potential parameter space for various SFSR inflation models 
\cite{Dai:2014jja,Munoz:2014eqa,Cook:2015vqa,Ueno:2016dim,Planck:2018jri,Hergt:2018ksk,Ellis:2021kad,Hergt:2022fxk}; these constraints 
have been obtained mainly using measurements of the scalar spectral index.

In this paper, we update the {\em Planck} inflationary analysis \cite{Planck:2018jri} including the latest BICEP/Keck data 
\cite{BICEP:2021xfz} focusing to a broader range of $\alpha$-attractor and D-brane inflationary models. We focus 
on the impact of different assumptions on the reheating phase and on the constraints on the reheating parameters 
derived from current cosmological observations.

This paper is organised as follows. In Section~\ref{sec:theory}, we review the computational method adopted to 
describe the primordial power spectra of scalar and tensor fluctuations for a given SFSR inflationary model. 
We also review the derivation of the number of $e$-folds, taking into account the uncertainties connected to an effective 
description of the reheating phase. 
We introduce the selection of inflationary models that we will study and their basic equations 
used for the analysis in Section~\ref{sec:models}. We present and discuss our results in Section~\ref{sec:results} 
for all the models analysed and for different parametrisation choice of reheating scenario.
We conclude in Section~\ref{sec:conclusions}.

\section{Slow-roll inflation predictions for the primordial power spectra} \label{sec:theory}
For a SFSR inflationary model, starting from the action given by
\begin{equation}
    S = \int \dd^4 x \sqrt{-g} \left[\frac{M_{\rm Pl}^2 R}{2} - \frac{1}{2}\partial_\mu\phi\partial^\mu\phi - V(\phi) \right] \,,
\end{equation}
the dynamical equations for the background, namely the Friedmann equation and the Klein-Gordon equation, are respectively 
\begin{subequations}
    \begin{align}
        &H^2 = \frac{1}{3 M_{\rm Pl}^2}\left(\frac{\dot{\phi}}{2} + V\right) \,,\\
        &\ddot{\phi} + 3H\dot{\phi} + V_\phi = 0 \,,
    \end{align}
\end{subequations}
where $V_\phi \equiv \dd V / \dd \phi$ and the background metric is the spatially-flat Friedmann-Robertson-Walker one given by 
$\dd s^2 = -\dd t^2 + a^2(t)\dd x^2 = a^2(t)(-\dd \tau^2 + \dd x^2)$. 
$M_{\rm Pl} \equiv (8\pi G)^{-1/2}$ is the reduced Planck mass. At leading order in perturbations, the 
equation of motion in term of gauge-invariant quantity $v_{\bf k}$ in Fourier space is given by \cite{Mukhanov:1981xt}
\begin{equation}
    v_{\bf k}'' + \left[k^2 - \frac{(a\sqrt{\epsilon_1})''}{a\sqrt{\epsilon_1}}\right]v_{\bf k} = 0
\end{equation}
for scalar perturbations where the scalar gauge-invariant quantity $v$, defined has $v_{\bf k} \equiv z\mathcal{R}_{\bf k}$, is related to the scalar metric perturbations through the gauge-invariant curvature perturbation variable $\mathcal{R}$ \cite{1992PhR...215..203M}. For tensor perturbations, we have 
\begin{equation}
    v_{\bf k}'' + \left[k^2 - \frac{a''}{a}\right]v_{\bf k} = 0 \,.
\end{equation}
where the tensor perturbations $h$ is already gauge invariant and therefore represent a physical degree of freedom. $v$ is defined as $v_{\bf k} \equiv a h_{\bf k}$.

Primordial power spectra (PPS) of scalar and tensor cosmological fluctuations can be described with an analytic 
perturbative expansion. The result is an unified framework to connect the predictions for hundreds of slow-roll 
inflationary models to cosmological observations \cite{Martin:2013tda,Martin:2024qnn}.

The method, developed in Ref.~\cite{Starobinsky:1979ty} for tensor perturbations and in 
Refs.~\cite{Mukhanov:1985rz,Mukhanov:1988jd} for scalar perturbations, has been improved and extended including 
higher-order corrections at next-to-leading order (NLO) in 
Refs.~\cite{Stewart:1993bc,Liddle:1994dx,Gong:2001he,Schwarz:2001vv,Leach:2002ar},  
and next-next-to-leading (NNLO) in Refs.~\cite{Auclair:2022yxs,Bianchi:2024qyp,Ballardini:2024irx}. Using the calculation based on 
the Green’s function method \cite{Gong:2001he}, the PPS for scalar (S) and tensor (T) perturbations can be expanded 
in terms of $\ln k$, around a particular reference scale $k_*$ up to NLO \cite{Leach:2002ar}, giving
\begin{equation}
    \ln \frac{P_{\rm X}(k)}{P_{\rm X0}(k_*)} = b_{\rm X0} + b_{\rm X1}\ln \left(\frac{k}{k_*}\right) 
    + \frac{1}{2}b_{\rm X2}\ln^2 \left(\frac{k}{k_*}\right) + \dots \,,
\end{equation}
where ${\rm X=\{S,\,T\}}$ and the normalisation of the PPS are given by
\begin{equation}
    P_{\rm S0} = \frac{H^2_*}{8\pi^2 M_{\rm Pl}^2 \epsilon_1} \,,\qquad P_{\rm T0} = \frac{2H^2_*}{\pi^2 M_{\rm Pl}^2} \,.
\end{equation}

The expansion coefficients, up to NLO of Hubble flow functions (HFF) 
$\epsilon_n$, are given by
\begin{subequations} \label{eqn:bs}
    \begin{align}
    b_{\rm S0} = &-2(1-\alpha)\epsilon_1+\alpha\epsilon_2+\left(2\alpha+\frac{\pi^2}{2}-5\right)\epsilon_1^2 \notag\\
            &+\left(-\alpha^2+3\alpha+\frac{7\pi^2}{12}-6\right)\epsilon_1\epsilon_2+\left(\frac{\pi^2}{8}-1\right)\epsilon_2^2 \notag\\
            &+\left(-\frac{\alpha^2}{2}+\frac{\pi^2}{24}\right)\epsilon_2\epsilon_3 \,,\\
    b_{\rm S1} = &-2\epsilon_1-\epsilon_2-2\epsilon_1^2-(3-2\alpha)\epsilon_1\epsilon_2-\alpha\epsilon_2\epsilon_3 \,,\\
    b_{\rm S2} = &-2\epsilon_1\epsilon_2-\epsilon_2\epsilon_3 \,,
\end{align}
\end{subequations}
for scalar perturbations, and
\begin{subequations} \label{eqn:bt}
\begin{align} 
    b_{\rm T0} = &-2(1-\alpha)\epsilon_1+\left(2\alpha+\frac{\pi^2}{2}-\frac{10}{2}\right)\epsilon_1^2 \notag\\
            &+\left(-\alpha^2+2\alpha+\frac{\pi^2}{12}-2\right)\epsilon_1\epsilon_2 \,,\\
    b_{\rm T1} = &-2\epsilon_1-2\epsilon_1^2-2(1-\alpha)\epsilon_1\epsilon_2 \,,\\
    b_{\rm T2} = &-2\epsilon_1\epsilon_2 \,,
\end{align}
\end{subequations}
for tensor perturbations. Here $\alpha \equiv \gamma_{\rm E} + \ln (2) - 2 \approx 0.7296$ and $\gamma_{\rm E}$ is 
the Euler-Mascheroni constant. The needed HFF to describe the NLO expansion are
\begin{subequations} \label{eqn:eps}
\begin{align}
    \epsilon_1 &= 2M_{\rm Pl}^2\left(\frac{H'}{H}\right)^2 \,,\\
    \epsilon_2 &= 4M_{\rm Pl}^2\left[\left(\frac{H'}{H}\right)^2 - \frac{H''}{H}\right] \,,\\
    \epsilon_3 &= 2M_{\rm Pl}^2\left[2\left(\frac{H'}{H}\right)^2+\frac{H'''}{H'}-3\frac{H''}{H}\right]\left(1-\frac{HH''}{H'^2}\right)^{-1} \,,
\end{align}
\end{subequations}
where a prime $'$ denotes derivative with respect to conformal time $\tau$. HFF can be calculated directly from 
a given single-field potential \cite{Liddle:1994dx} as
\begin{subequations} \label{eqn:epsv}
\begin{align}
    \epsilon_1 &\simeq \frac{M_{\rm Pl}^2}{2}\left(\frac{V_\phi}{V}\right)^2 \,,\\
    \epsilon_2 &\simeq 2M_{\rm Pl}^2\left[\left(\frac{V_\phi}{V}\right)^2 - \frac{V_{\phi\phi}}{V}\right] \,,\\
    \epsilon_2\epsilon_3 &\simeq 2M_{\rm Pl}^4\left[\frac{V_{\phi\phi\phi}V_\phi}{V^2} - 3\frac{V_{\phi\phi}}{V}\left(\frac{V_\phi}{V}\right)^2 + 2\left(\frac{V_{\phi}}{V}\right)^4\right] \,.
\end{align}
\end{subequations}

\subsection{Effective description of the reheating phase}
In order to derive accurate predictions, we need to calculate the number of $e$-folds from the time that a given 
perturbation scale leaves the horizon until the end of inflation. This requires knowledge about the end of inflation, 
how the Universe reheats, and the post-inflationary evolution of the Universe, for a given model; see 
Refs.~\cite{Dodelson:2003vq,Liddle:2003as,Martin:2006rs}.

We can expand the definition of comoving Hubble scale $k = a_k H_k$ evaluated at the time of horizon exit such 
that
\begin{equation} \label{eqn:k}
    \frac{k}{a_0 H_0} = \frac{a_k}{a_{\rm end}}\frac{a_{\rm end}}{a_{\rm re}}\frac{a_{\rm re}}{a_{\rm eq}}\frac{H_k}{H_{\rm eq}}\frac{a_{\rm eq}H_{\rm eq}}{a_0H_0} \,.
\end{equation}
The number of $e$-folds between the time at which the comoving wavenumber $k$ crossing the comoving Hubble radius and 
the end of inflation is defined as $e^{N_k} \equiv a_{\rm end}/a_k$. The number of $e$-folds between the end of inflation 
and the beginning of the radiation-dominated phase, dubbed as {\it reheating phase}, is $e^{N_{\rm re}} \equiv a_{\rm re}/a_{\rm end}$. 
The duration of reheating can be described through an effective average equation of state
\begin{align} \label{eqn:Rre}
    \rho_{\rm re} =&\, \rho_{\rm end}e^{3\int\frac{{\rm d}a}{a}\left[1+w_{\rm re}(a)\right]} \notag\\
    =&\, \rho_{\rm end}e^{3\int_{N_{\rm end}}^{N}{\rm d}N'\left[1+w_{\rm re}(N')\right]} \notag\\
    =&\, \rho_{\rm end}e^{-3N_{\rm re}\left(1+\bar{w}_{\rm re}\right)} \,,
\end{align}
where the final energy density during reheating can be expressed as
\begin{equation} \label{eqn:rhore}
    \rho_{\rm re} = \frac{\pi^2}{30}g_{\rm re}T^4_{\rm re} \,,
\end{equation}
with $g_{\rm re}$ being the effective number of relativistic species upon thermalisation. Combining 
Eq.~\eqref{eqn:rhore} and Eq.~\eqref{eqn:Rre}, we can express the reheating temperature as
\begin{equation} \label{eqn:Tre}
    T_{\rm re} = \frac{30\rho_{\rm end}}{\pi^2g_{\rm re}}e^{-3N_{\rm re}(1+\bar{w}_{\rm re})} \,.
\end{equation}
The reheating temperature can be related to the CMB temperature today $T_\gamma$ assuming that the reheating 
entropy is conserved in the CMB and neutrino background today, that corresponds to entropy conservation 
${\rm d}(sa^3)=0$. This gives
\begin{equation} \label{eqn:ds0}
    a^3_{\rm re}g_{\rm s,\,re}T^3_{\rm re} = a^3_0\left(2+\frac{7}{8}2\frac{4}{11}N_{\rm eff}\right)T^3_\gamma \,,
\end{equation}
where $g_{\rm s,\,re}$ is the effective number of relativistic degrees of freedom for entropy at the end of reheating 
and we assumed that neutrino temperature today is given by $T_\nu = (4/11)^{1/3}T_\gamma$. 
$N_{\rm eff}$ is the effective number of neutrino families. Combining Eq.~\eqref{eqn:Tre} 
with Eq.~\eqref{eqn:ds0}, we obtain
\begin{equation}
    \frac{a_{\rm re}}{a_{\rm 0}} = \left(2+\frac{7}{11}N_{\rm eff}\right)^{1/3} T_\gamma \frac{g_{\rm re}^{1/4}}{g_{\rm s,\,re}^{1/3}}
    \left(\frac{\pi^2}{30\rho_{\rm end}}\right)^{1/4} e^{\frac{3}{4}N_{\rm re}\left(1+\bar{w}_{\rm re}\right)} \,.
\end{equation}
The last term of Eq.~\eqref{eqn:k} can be calculated from the Friedmann equation in the form
\begin{align}
    \frac{H_{\rm eq}}{H_0} &= \sqrt{(1+z_{\rm eq})^4\Omega_\gamma + (1+z_{\rm eq})^3\Omega_{\rm m} + \Omega_\Lambda} \notag\\
    &= \sqrt{2(1+z_{\rm eq})^3\Omega_{\rm m}} = 218.65\,(1+z_{\rm eq})\,\Omega_{\rm m}h \,,
\end{align}
where we assumed a spatially-flat background with $\Omega_k = 0$, neglected $\Omega_\Lambda$, and 
replaced
\begin{equation}
    1+z_{\rm eq} = \frac{\Omega_{\rm m} \rho_{\rm cr}}{\frac{\pi^2}{30}\left(2+\frac{7}{8}2\frac{4}{11}N_{\rm eff}\right)T^4_\gamma} \,,
\end{equation}
with $\rho_{\rm cr} \equiv 3H^2 M_{\rm Pl}^2$ the critical density. 
We obtain for the number of $e$-folds between horizon 
crossing and the end of inflation
\begin{equation} \label{eqn:Nk}
    N_k =\, 67.27 - \ln \left(\frac{k}{a_0 H_0}\right) 
    - \frac{1}{12} \ln\left(\frac{g_{\rm s,\,re}^4}{g_{\rm re}^3}\right) - \frac{1-3\bar{w}_{\rm re}}{4}N_{\rm re} 
    + \frac{1}{4} \ln\left(\frac{H_k^4}{\rho_{\rm end}}\right) \,.
\end{equation}
We split the last term on the right hand side separating the dependence from the end of inflation to the one at the 
horizon crossing as
\begin{equation}
    \frac{1}{4} \ln\left(\frac{H_k^4}{\rho_{\rm end}}\right) = \frac{1}{4} \ln\left(\frac{V_k}{3-\epsilon_{1,\,k}}\frac{3-\epsilon_{\rm 1,\,end}}{3V_{\rm end}}\right) + \frac{1}{4} \ln\left(8\pi^2A_{\rm s}\epsilon_{1,\,k}\right) \,,
\end{equation}
where $A_{\rm s} \equiv P_{\rm S0}(k_*)$. Inserting in Eq.~\eqref{eqn:Nk}, we obtain
\begin{align} \label{eqn:Nk2}
    N_k =&\, 66.72 - \ln \left(\frac{k}{a_0 H_0}\right) \notag\\
    &- \frac{1}{12} \ln\left(\frac{g_{\rm s,\,re}^4}{g_{\rm re}^3}\right) - \frac{1-3\bar{w}_{\rm re}}{4}N_{\rm re} \notag\\
    &+ \frac{1}{4} \ln\left(\frac{3V_k\epsilon_{1,\,k}}{3-\epsilon_{1,\,k}}\frac{3-\epsilon_{\rm 1,\,end}}{V_{\rm end}}\right) + \frac{1}{4} \ln\left(8\pi^2A_{\rm s}\right) \,.
\end{align}
 Using Eq.~\eqref{eqn:Rre}, we can rewrite the number of $e$-folds during reheating as
\begin{equation} \label{eqn:Nre}
    N_{\rm re} = -\frac{1}{3+3\bar{w}_{\rm re}}\ln\left(\frac{\rho_{\rm re}}{\rho_{\rm end}}\right) \,.
\end{equation}

Eq.~\eqref{eqn:Nk2} allows for an accurate value for the number of $e$-folds between horizon crossing and the end of 
inflation. We fix the values of the cosmological parameters to $N_{\rm eff} = 3.044$ 
\cite{2020JCAP...08..012A,Froustey:2020mcq,2021JCAP...04..073B}, 
$T_\gamma = 2.7255\,{\rm K}$ \cite{Fixsen:2009ug}, $H_0 = 67.36\,{\rm km}\,{\rm s}^{-1}\,{\rm Mpc}^{-1}$, 
and $\Omega_{\rm m} = 0.3153$ \cite{Planck:2018vyg}, the pivot scale to $k_* = 0.05\,{\rm Mpc}^{-1}$. 
It is reasonable to assume all the particles to be in thermal equilibrium in the early Universe, that corresponds 
to $g_{\rm re} = g_{\rm s,\,re}$, and we fix $g_{\rm re} = 106.75$ to the Standard Model (SM) prediction even if a larger 
value for $g_{\rm re}$ might arise at high energies in beyond SM theories.

Finally, the most convenient way to write the number of $e$-folds is
\begin{align} \label{eqn:Nk3}
    N_{0.05} =&\, 60.9 + \frac{1-3\bar{w}_{\rm re}}{12+12\bar{w}_{\rm re}} \ln\left(\frac{\rho_{\rm re}}{M_{\rm Pl}^4}\right) 
    + \frac{1+3\bar{w}_{\rm re}}{6+6\bar{w}_{\rm re}} \ln\left(8\pi^2A_{\rm s}\right) \notag\\
    &+ \frac{1}{3+3\bar{w}_{\rm re}} \ln\left[\frac{V_*}{V_{\rm end}}\frac{2}{3-\epsilon_{1,\,*}}(3\epsilon_{1,\,*})^\frac{1+3\bar{w}_{\rm re}}{2}\right] \,,
\end{align}
where we consider as parameters $\ln(\rho_{\rm re}/M_{\rm Pl}^4)$, $A_{\rm s}$, $\bar{w}_{\rm re}$ 
(hereafter we will simple use $w_{\rm re}$ for the effective average equation of state during reheating), and 
$\epsilon_{1,\,{\rm end}} = 1$.

\section{Inflationary models and slow-roll dynamics} \label{sec:models}
For single-field inflationary models with a standard kinetic term and a potential $V(\phi)$, we can calculate analytically  
the PPS of scalar and tensor fluctuations given a shape of inflationary potential 
combining Eq.~\eqref{eqn:bs} and Eq.~\eqref{eqn:bt} with Eq.~\eqref{eqn:epsv}. Finally, in order to have expressions 
in terms of the number of $e$-folds rather than values of the scalar field, we need to solve the expression of the classical 
inflationary trajectory
\begin{equation} \label{eqn:trajectory}
    N_* \equiv N_{\rm end} - N_* = -\frac{1}{M_{\rm Pl}^2} \int_{\phi_*}^{\phi_{\rm end}} \dd\phi\, \frac{V}{V_\phi} \,.
\end{equation}

In the following, we will provide the expression of the inflationary trajectory $\phi(N)$ and for the value of the 
field at which inflation end $\phi_{\rm end}$, such that $\epsilon_{1,\,{\rm end}} \equiv \epsilon_1(\phi_{\rm end}) = 1$, 
for each of the inflationary models studied.

\subsection{$R + R^2$ inflation}
The first inflationary model proposed in Ref.~\cite{Starobinsky:1980te} is based on higher-order gravitational terms 
as 
\begin{equation}
    \frac{{\cal L}}{\sqrt{-g}} = \frac{1}{2}\left(R + \frac{R^2}{6M^2}\right) \,.
\end{equation}
In the conformally-related Einstein frame (EF) \cite{Maeda:1988ab}, it corresponds to a scalar field $\phi$ with potential
\begin{equation} \label{eqn:V_R2}
    V_{\rm R+R^2}(\phi) = V_0 \left(1 - e^{-\sqrt{\frac{2}{3}}\phi}\right)^2 \,.
\end{equation}
The condition $\epsilon_{1,\,{\rm end}} = 1$ occurs for
\begin{equation}
    \phi_{\rm end} = \sqrt{\frac{3}{2}} \ln \left(1 + \frac{2}{\sqrt{3}}\right) \,,
\end{equation}
and the slow-roll trajectory can be expressed inverting Eq.~\eqref{eqn:trajectory} as 
\begin{align} \label{eqn:phi_R2}
    \phi_* =&\, \sqrt{\frac{3}{2}}\left\{ -\frac{4}{3}N_* - \left(1 + \frac{2}{\sqrt{3}}\right) + \ln \left(1 + \frac{2}{\sqrt{3}}\right) \right.\notag\\
    &\left.- W_{-1}\left[-e^{-\frac{4}{3}N_* - \left(1 + \frac{2}{\sqrt{3}}\right) + \ln \left(1 + \frac{2}{\sqrt{3}}\right)}\right]\right\} \,,
\end{align}
where $W_{-1}$ is the Lambert function in the $-1$-branch. 

We can now derive predictions for the scalar spectral index $n_{\rm s}$ and the tensor-to-scalar ratio $r$ calculating 
Eq.~\eqref{eqn:epsv} with Eq.~\eqref{eqn:V_R2} and Eq.~\eqref{eqn:phi_R2}. The leading-order predictions in the limit 
$N \gg 1$ \cite{Mukhanov:1981xt,Starobinsky:1983zz} are
\begin{equation} \label{eqs:approx_R2}
    n_{\rm s} \approx 1-\frac{2}{N} \,,\qquad r \approx \frac{12}{N^2} \,.
\end{equation}
The same predictions come from a scalar field model with $V(\phi) = \lambda\phi^4/4$ at large values of $\phi$ and a 
large non-minimal coupling to gravity $\xi R\phi^2$, including the Higgs inflation model \cite{Bezrukov:2007ep}.

Solving Eq.~\eqref{eqn:Nk3} for the number of $e$-folds at horizon crossing one finds $N_{0.05} \simeq 54.8$ for 
instantaneous reheating, that corresponds to $N_{\rm re} = 0$, and $N_{0.05} \simeq 50.3$ for 
$T_{\rm re} = 3.1\times 10^9\,{\rm GeV}$ \cite{Gorbunov:2010bn,Bezrukov:2011gp} assuming that the Universe is in a 
matter-dominated stage while the scalaron decays into the SM Higgs bosons \cite{Starobinsky:1980te,Vilenkin:1985md}. 
It is important to stress that while for these values leading-order predictions \eqref{eqs:approx_R2} lead to
\begin{align}
    &n_{\rm s}(N_{0.05} \simeq 54.8) = 0.9635 \,,\qquad r(N_{0.05} \simeq 54.8) = 0.0040 \,,\\
    &n_{\rm s}(N_{0.05} \simeq 50.3) = 0.9602 \,,\qquad r(N_{0.05} \simeq 50.3) = 0.0047 \,,
\end{align}
using second-order slow-roll analytic predictions gives
\begin{subequations} \label{eqn:R2_pred}
\begin{align}
    &n_{\rm s}(N_{0.05} \simeq 54.8) = 0.9653 \,,\qquad r(N_{0.05} \simeq 54.8) = 0.0034 \,,\\
    &n_{\rm s}(N_{0.05} \simeq 50.3) = 0.9623 \,,\qquad r(N_{0.05} \simeq 50.3) = 0.0040 \,,
\end{align}
\end{subequations}
resulting in a 15\% difference.

\subsection{Cosmological attractors}
Many cosmological attractor models have been proposed to make the inflationary predictions of simple scalar fields 
compatible with cosmological data generalising their kinetic term. Two simple examples of $\alpha$-attractors 
\cite{Kallosh:2013tua,Ferrara:2013rsa,Kallosh:2013yoa,Kallosh:2014rga,Kallosh:2014laa,Galante:2014ifa} 
are given by
\begin{equation}
    \frac{\cal L}{\sqrt{-g}} = \frac{R}{2} 
    - \frac{1}{2}\frac{\left(\partial_\mu \phi\right)^2}{\left(1-\frac{\phi^2}{6\alpha}\right)^2} - V(\phi) \,,
\end{equation}
the so called T-model, and the E-model
\begin{equation}
    \frac{\cal L}{\sqrt{-g}} = \frac{R}{2} 
    - \frac{3\alpha}{4}\frac{\left(\partial_\mu \phi\right)^2}{\phi^2} - V(\phi) \,.
\end{equation}
$\alpha$-attractors represent a special class of pole inflation models \cite{Galante:2014ifa}, described by the equation
\begin{equation}
    \frac{\cal L}{\sqrt{-g}} = \frac{R}{2} 
    - \frac{a_q}{2}\frac{\left(\partial_\mu \phi\right)^2}{\phi^q} - V(\phi) \,,
\end{equation}
where $q=2$, $a_2=3\alpha/2$ reduces to the E-models of $\alpha$-attractors.\footnote{See Ref.~\cite{Kallosh:2013hoa} for 
previously introduced T-model and the E-model conformal attractors.}
The origin of the pole in the kinetic term can be explained in the context of hyperbolic geometry in supergravity and string 
theory \cite{Carrasco:2015uma}, or related to a non-minimal coupling of the inflaton field to gravity \cite{Kallosh:2013pby,Galante:2014ifa,Kallosh:2013tua}. 
Using a canonical normalised scalar field $\varphi$, it is possible to rewrite the the theory as
\begin{equation}
    \frac{\cal L}{\sqrt{-g}} = \frac{R}{2} 
    - \frac{\left(\partial_\mu \varphi\right)^2}{2} - V(\varphi) \,,
\end{equation}
where the potential and its derivatives are not singular. For the simplest case $V(\phi) \propto \phi^{2n}$, in terms of 
the canonical variables, we have 
\begin{equation} \label{eqn:VT}
    V_{\rm T-model}(\varphi) = V_0 \tanh^{2n}\left({\frac{\varphi}{\sqrt{6\alpha}}}\right) \,,
\end{equation}
and 
\begin{equation} \label{eqn:VE}
    V_{\rm E-model}(\varphi) = V_0 \left(1 - e^{-\sqrt{\frac{2}{3\alpha}}\varphi}\right)^{2n} \,.
\end{equation}
In case of $n = 1$, Eq.~\eqref{eqn:VE} corresponds to the EF potential in the $R+R^2$ inflation \cite{Starobinsky:1979ty} 
and Higgs inflation \cite{Bezrukov:2007ep} for $\alpha = 1$, and Goncharov-Linde (GL) model of chaotic inflation in supergravity 
\cite{Goncharov:1983mw} for $\alpha = 1/9$. The case of $\alpha = 2,\,n=1/2$ corresponds to fibre inflation \cite{Cicoli:2008gp,Kallosh:2017wku}.
Values corresponding to $3\alpha = 7,\, 6,\, 5,\, 4,\, 3,\, 2,\, 1$ are associated to supergravity models and are usually called 
Poincar\'e disk models \cite{Ferrara:2016fwe,Kallosh:2019hzo}.

The condition $\epsilon_{1,\,{\rm end}} = 1$ for T-model and E-model occurs for
\begin{align}
    \varphi_{\rm end}^{\rm T} &= \sqrt{\frac{3 \alpha}{2}} \sinh ^{-1}\left(\frac{2n}{\sqrt{3\alpha}}\right) \,,\\
    \varphi_{\rm end}^{\rm E} &= \sqrt{\frac{3 \alpha }{2}} \ln \left(\frac{2 n}{\sqrt{3\alpha}}+1\right)  \,,
\end{align}
and the slow-roll trajectories can be expressed as 
\begin{align}
    \varphi_*^{\rm T} =&\, \sqrt{\frac{3\alpha}{2}} {\rm sech}^{-1}\left(\frac{3 \alpha}{\alpha \sqrt{\frac{12n^2}{\alpha}+9} + 4nN_*}\right) \,,\\
    \varphi_*^{\rm E} =&\, \sqrt{\frac{3\alpha}{2}}\left\{ -\frac{4n}{3\alpha}N_* 
    -\left(1 + \frac{2n}{\sqrt{3\alpha}}\right) + \ln \left(1 + \frac{2n}{\sqrt{3\alpha}}\right)\right. \notag\\
    &\left.- W_{-1}\left[ -e^{-\frac{4n}{3\alpha}N_* 
    -\left(1 + \frac{2n}{\sqrt{3\alpha}}\right) + \ln \left(1 + \frac{2n}{\sqrt{3\alpha}}\right)} \right]\right\} \,.
\end{align}

The predictions of the T-models \eqref{eqn:VT} coincide with the ones of the E-models \eqref{eqn:VE} in the limits 
for $\alpha \to 0$ and $\alpha \to \infty$. For $\alpha \gg 1$, the model predictions correspond to the ones for 
large-field chaotic models $V(\varphi) \propto \varphi^{2n}$ \cite{Linde:1983gd}, while for $\alpha \ll 1$ to
\begin{equation}
    n_{\rm s} \approx 1-\frac{2}{N}\,,\qquad r \approx \frac{12\alpha}{N^2}
\end{equation}
for both T-models and E-models and for all values $n$. For $\alpha < 1$, $\alpha$-attractors predict a value of the 
tensor-to-scalar ratio smaller than the one predicted in $R+R^2$ inflation.

\subsection{D-brane inflation}
String theory D-brane inflation models \cite{Kachru:2003sx,Dvali:2001fw,Burgess:2001fx,Garcia-Bellido:2001lbk} correspond to 
Dp-brane-$\overline{\rm Dp}$-brane interaction where the inflationary potentials have the form
\begin{equation} \label{eqn:V_BI}
    V_{\rm BI}(\phi) = V_0 \left[1 - \left(\frac{m}{\phi}\right)^{7-p} + \dots \right] \,,
\end{equation}
in brane inflation, and 
\begin{equation} \label{eqn:V_KKLT}
    V_{\rm KKLTI}(\phi) = V_0 \left[1 + \left(\frac{m}{\phi}\right)^{7-p} \right]^{-1} \,,
\end{equation}
in Kachru-Kallosh-Linde-Trivedi (KKLT) inflation \cite{Kachru:2003aw}. 
Models well compatible with cosmological observations are the inverse quadratic (D5-$\overline{\rm D5}$) and inverse quartic 
(D3-$\overline{\rm D3}$) \cite{Kallosh:2018zsi,Kallosh:2019jnl}, both associated with type IIB string theory and possible moduli 
stabilisation due to KKLT \cite{Kachru:2003aw} and LVT \cite{Balasubramanian:2005zx,Conlon:2005ki} construction.
In addition, the inverse linear case with D6-$\overline{\rm D6}$ potential in type IIA string theory \cite{Kallosh:2018nrk,Blaback:2018hdo}.
In this case 
\begin{equation} \label{eqn:V_BI1}
    V_{\rm BI}(\phi) = V_0 \left(1 - \frac{m}{|\phi|} + \dots \right) \,,
\end{equation}
and 
\begin{equation} \label{eqn:V_KKLT1}
    V_{\rm KKLTI}(\phi) = V_0 \left(1 + \frac{m}{|\phi|} \right)^{-1} \,,
\end{equation}
where $\phi$ is a distance in the moduli space.
Brane inflation potential \eqref{eqn:V_BI} is unbounded from below and it requires a consistent generalisation, in particular for 
the parameter space region allowed by data, that is $\phi < m$. The predictions of \eqref{eqn:V_BI} coincide with 
the one of \eqref{eqn:V_KKLT} in the limit for $\phi < m$.

For the KKLT potential there is no generic solution for $\phi_{\rm end}$. Analytical solutions can be find for integer values 
of $p$ or for the limits $\phi \ll m$ and $\phi \gg m$.

Similarly to $\alpha$-attractors, KKLT models have universal predictions for $m \lesssim 1$ and small $r$. In this limit, we have 
\begin{equation}
    ^4n_{\rm s} \approx 1 - \frac{5}{3N} \,,\qquad ^2n_{\rm s} \approx 1 - \frac{3}{2N} \,,\qquad ^1n_{\rm s} \approx 1 - \frac{4}{3N} 
\end{equation}
corresponding to the same predictions of $n_{\rm s}$ for $V(\phi) \propto \phi^{2n}$ with $n = \frac{7-p}{9-p}$, and
\begin{equation}
    ^4r \approx \frac{4m^{4/3}}{(3N)^{5/3}} \,,\qquad ^2r \approx \frac{12m}{N^2} \,,\qquad ^1r \approx \frac{8m^{2/3}}{(3N)^{3/4}} \,,
\end{equation}
where the prefix refers to the value $7 - p$.

\section{Analysis and results} \label{sec:results}
We use {\tt CosmoMC} \cite{Lewis:2013hha} connected to our modified version of the code {\tt CAMB} 
\cite{Lewis:1999bs,Howlett:2012mh} sampled with the nested sampling code {\tt PolyChord} 
\cite{Handley:2015fda,Handley:2015vkr}, which allow to obtain simultaneously the log-evidence. 
Mean values and uncertainties on the parameters, as well as the posterior 
distributions plotted, have been generated using {\tt GetDist} \cite{Lewis:2019xzd}. 
For the computation of the Kullback-Leibler (KL) divergence, we rely on {\tt anesthetic} \cite{Handley:2019mfs}.

We use {\em Planck} temperature, polarisation, and lensing 2018 legacy PR3 data \cite{Planck:2019nip} (hereafter P18). 
Low-multipole data for $\ell < 30$ consists to the {\tt commander} likelihood for temperature and {\tt SimAll} 
for the E-mode polarisation. On high multipoles $\ell \geq 30$, we use the {\tt Plik} likelihood including 
CMB temperature up to $\ell_{\rm max} = 2508$, E-mode polarisation and temperature-polarisation cross correlation 
up to $\ell_{\rm max} = 1996$.
We include B-mode polarisation spectrum for $20 < \ell < 330$ from BICEP2, Keck Array, and BICEP3 
observations up to 2018 \cite{BICEP:2021xfz} (hereafter BK18). 
Additionally, we include measurements of baryon acoustic oscillations (BAO) and redshift space distortions 
(RSD) at low redshift $0.07 < z < 0.2$ from SDSS-I and -II sample as {\em Main Galaxy Sample} (MGS), BOSS DR12 
galaxies over the redshift interval $0.2 < z < 0.6$, eBOSS luminous red galaxies (LRG) and quasars $0.6 < z < 2.2$, 
and Lyman-$\alpha$ forest samples $1.8 < z < 3.5$ \cite{eBOSS:2020yzd}.
We also include the Pantheon catalogue of uncalibrated Type Ia Supernovae (SNe) over the redshift range $0.01 < z < 2.3$ 
\cite{Pan-STARRS1:2017jku}.

In addition to the inflationary parameters discussed in the previous section, we vary the standard cosmological parameters 
$\omega_{\rm b},\, \omega_{\rm c},\, \theta_{\rm MC},\, \tau,\, A_{\rm s}$, as well as nuisance and foreground parameters. 
As baseline, we allow the reheating phase to last down to $\rho_{\rm re}^{1/4} = 1\,{\rm TeV}$ maximum and to happen in a 
matter-dominated phase, corresponding to $w_{\rm re} = 0$.\footnote{For inflationary potentials that can be approximated to 
$V(\phi) \propto \phi^{2n}$ around their minima, after inflation, the homogeneous inflaton field oscillates initially with 
average equation of state given by $\bar{w}_{\rm hom} = (n - 1)/(n + 1)$ \cite{Turner:1983he}.} 
We will present in the next section results for different assumptions 
on the reheating phase. Prior ranges on the standard and inflationary parameters are collected on Table~\ref{tab:prior_param}. 
We report a rough estimation of the allowed prior ranges for some derived parameters in Table~\ref{tab:prior_inf}. 
Note that $\alpha$-attractors can easily describe any value of $r \ll 1$ without spoiling the predictions on the scalar spectral 
index. In Table~\ref{tab:prior_inf}, the lower value for the tensor-to-scalar ratio reflects the prior range adopted on the 
parameter $\alpha$, sufficiently large for the sensitivity of current CMB measurements.

\begin{table*}
\centering
\begin{tabular}{l|c}
\hline
\hline
Parameter & Uniform prior \\
\hline
$\omega_{\rm b} \equiv \Omega_{\rm b} h^2$                & [0.019, 0.025] \\
$\omega_{\rm c} \equiv \Omega_{\rm c} h^2$                & [0.095, 0.145] \\
$100\theta_{\rm MC}$                   & [1.03, 1.05]   \\
$\tau$                              & [0.01, 0.4]    \\
$\ln\left(10^{10}A_{\rm s}\right)$  & [2.5, 3.7]     \\
$\ln\left(\rho_{\rm re}/M_{\rm Pl}^4\right)$  & [$\ln\left(1\,{\rm TeV}/M_{\rm Pl}^4\right)$, $\ln\left(\rho_{\rm end}/M_{\rm Pl}^4\right)$]     \\
$\log \alpha^{\rm T}$               & [-2, 4]        \\
$\log \alpha^{\rm E}$               & [-2, 4]        \\
$\log m$                          & [-4, 4]        \\
\hline
\end{tabular}
\caption{Prior ranges for cosmological parameters used in the Bayesian comparison of inflationary models.\label{tab:prior_param}}
\end{table*}

\begin{table*}
\centering
\begin{tabular}{l|ccc}
\hline
\hline
Model                        & $N_{0.05}$      & $n_{\rm s,\,0.05}$         & $r_{0.05}$ \\
\hline
$R+R^2$                 & $[45,\, 55]$      & $[0.958,\, 0.966]$          & $[0.0034,\, 0.0049]$ \\
T-model $n=1/2$    & $[44,\, 56]$      & $[0.955,\, 0.973]$          & $[3\times 10^{-5},\, 0.086]$ \\
T-model $n=2/3$    & $[44,\, 56]$      & $[0.955,\, 0.971]$          & $[4\times 10^{-5},\, 0.11]$ \\
T-model $n=1$      & $[44,\, 57]$      & $[0.955,\, 0.964]$          & $[4\times 10^{-5},\, 0.17]$ \\
T-model $n=3/2$    & $[44,\, 57]$      & $[0.947,\, 0.964]$          & $[4\times 10^{-5},\, 0.25]$ \\
GL                 & $[44,\, 54]$      & $[0.957,\, 0.964]$          & $[4.3\times 10^{-4},\, 6.2\times 10^{-4}]$ \\
Poincar\'e         & $[45,\, 55]$      & $[0.959,\, 0.966]$          & $[0.007,\, 0.010]$ \\
E-model $n=1/2$    & $[44,\, 56]$      & $[0.955,\, 0.973]$          & $[4\times 10^{-5},\, 0.080]$ \\
E-model $n=2/3$    & $[44,\, 56]$      & $[0.955,\, 0.971]$          & $[4\times 10^{-5},\, 0.11]$ \\
E-model $n=1$      & $[44,\, 57]$      & $[0.949,\, 0.964]$          & $[4\times 10^{-5},\, 0.16]$ \\
E-model $n=3/2$    & $[44,\, 57]$      & $[0.949,\, 0.963]$          & $[4\times 10^{-5},\, 0.23]$ \\
KKLT $p=3$              & $[42,\, 58]$      & $[0.937,\, 0.968]$          & $[4\times 10^{-9},\, 0.32]$ \\
KKLT $p=5$              & $[43,\, 57]$      & $[0.957,\, 0.972]$          & $[3\times 10^{-7},\, 0.17]$ \\
KKLT $p=6$              & $[44,\, 56]$      & $[0.967,\, 0.976]$          & $[2\times 10^{-5},\, 0.09]$ \\
\hline
\end{tabular}
\caption{Allowed ranges for some derived inflationary parameters taking into account Eq.~\eqref{eqn:Nk3} and propagating 
the dependence on the variation of $\alpha$ for $\alpha$-attractors and $m$ for KKLT inflation, 
respectively.\label{tab:prior_inf}}
\end{table*}

\begin{figure}
\centering
\includegraphics[width=0.32\textwidth]{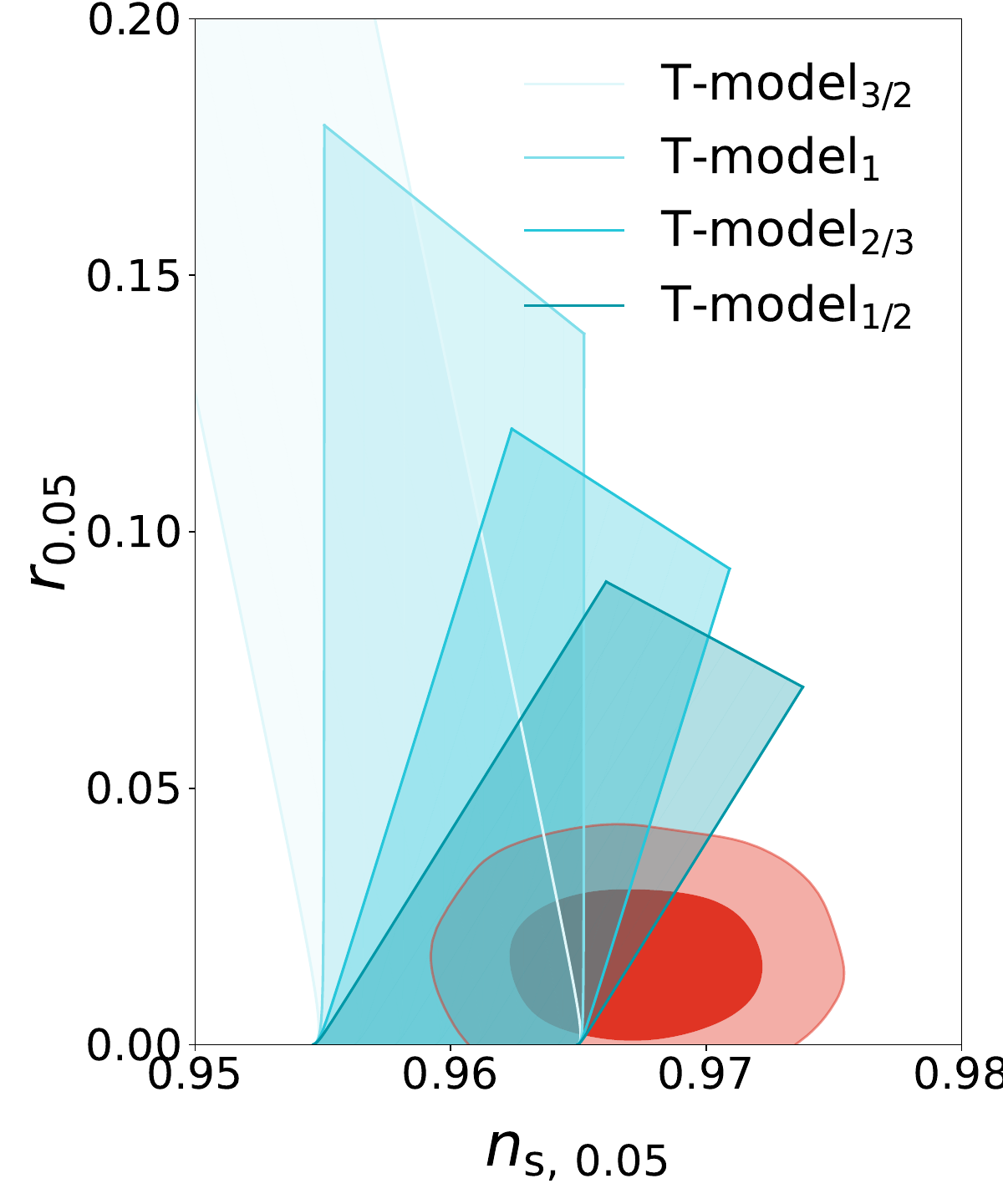}
\includegraphics[width=0.32\textwidth]{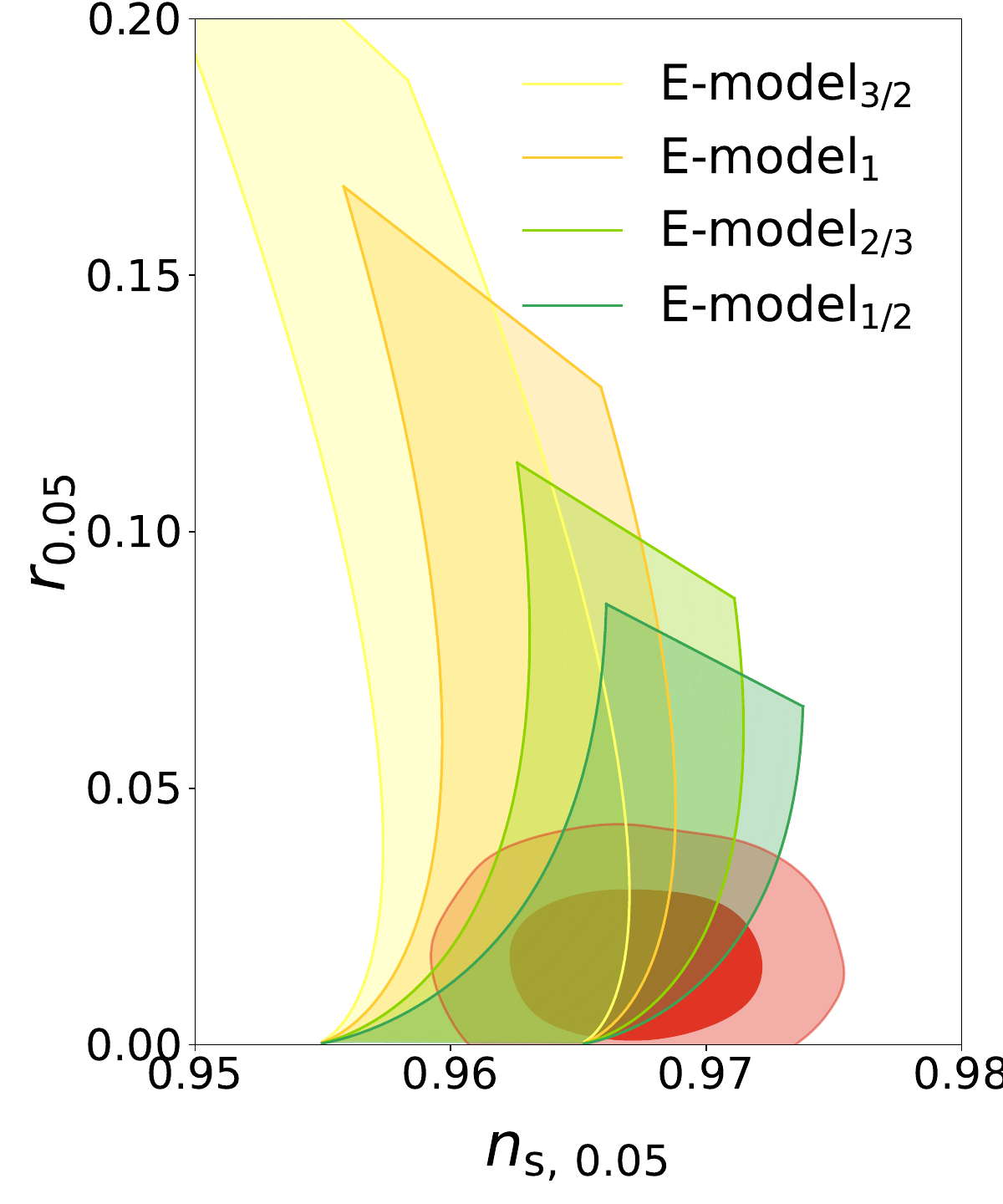}
\includegraphics[width=0.32\textwidth]{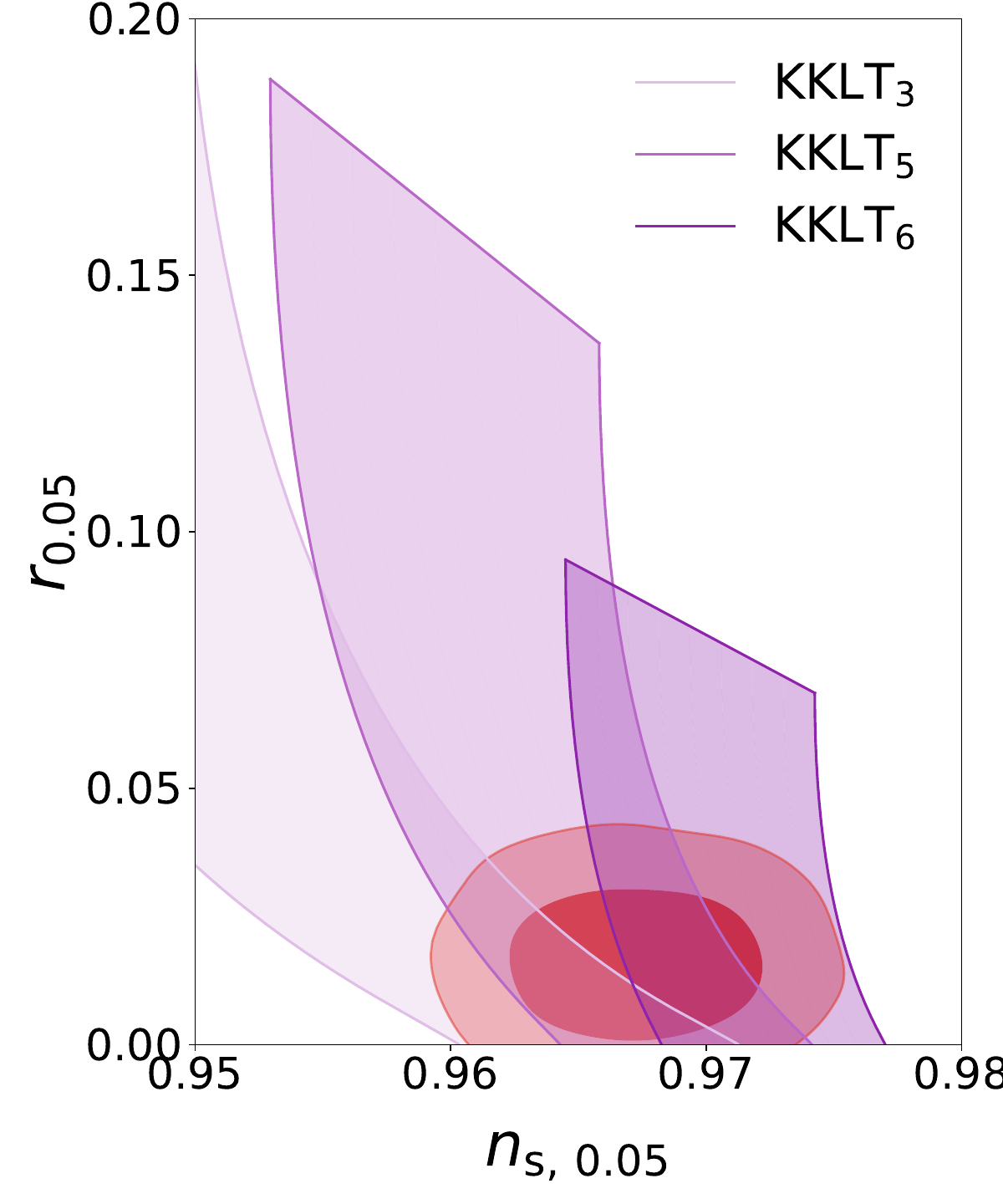}
\caption{Marginalised joint confidence contours for the scalar spectral index $n_{\rm s,\,0.05}$ and tensor-to-scalar ratio 
$r_{0.05}$ for the $\Lambda$CDM+$r$ model at 68\% CL and 95\% CL compared to the theoretical predictions 
for T-model $\alpha$-attractor (left panel), E-model $\alpha$-attractor (central panel), and KKLT (right panel) inflation.
\label{fig:nsr_theory}}
\end{figure}
Theoretical predictions for the scalar spectral index and the tensor-to-scalar ratio at $k_* = 0.05\,{\rm Mpc}^{-1}$ 
are shown in Fig.~\ref{fig:nsr_theory} for the range of parameters considered in the analysis given in Table~\ref{tab:prior_param} and in Table~\ref{tab:prior_inf}.

\subsection{Model comparison for inflationary models}
We perform a Bayesian analysis of the combination of datasets described above given the model parameters, including the 
reheating uncertainties. Here we sample directly on the inflationary parameters rather than sampling on slow-roll parameters 
$\epsilon_n$ or the phenomenological ones ($n_{\rm s},\, \alpha_{\rm s},\, n_{\rm t}, \dots$) to described the shape of the PPS. 
We have zero extra parameters for $R+R^2$ inflation (synonymous of Starobinsky inflation) and for GL inflation 
(E-model with $n = 1$ and $\alpha^{\rm E} = 1/9$)  
while one extra parameter for the other inflationary models considered. We have $\alpha^{\rm T}_n$ for T-model $\alpha$-attractor, 
$\alpha^{\rm E}_n$ for E-model $\alpha$-attractor, and $m_p$ for KKLT inflation. 

All the results are presented in comparison to the spatially-flat $\Lambda$CDM+$r$ model to highlight the differences on the posterior 
distributions of $n_{\rm s}$ and $r$, that are derived parameters in our cases. This shows that the results are 
dominated by theoretical prior knowledge on the models injected in the analysis. 

\begin{figure}
\centering
\includegraphics[width=0.49\textwidth]{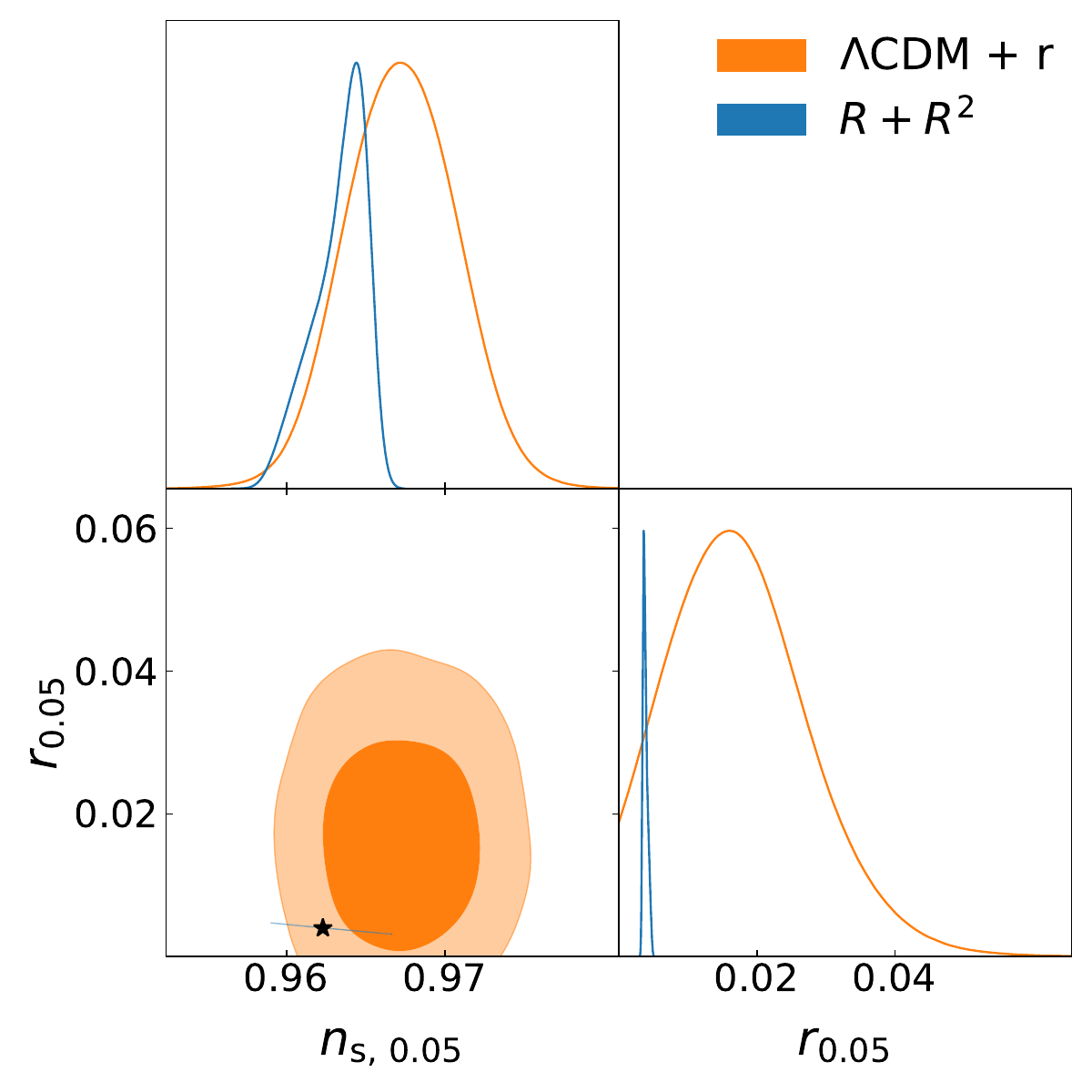}
\includegraphics[width=0.49\textwidth]{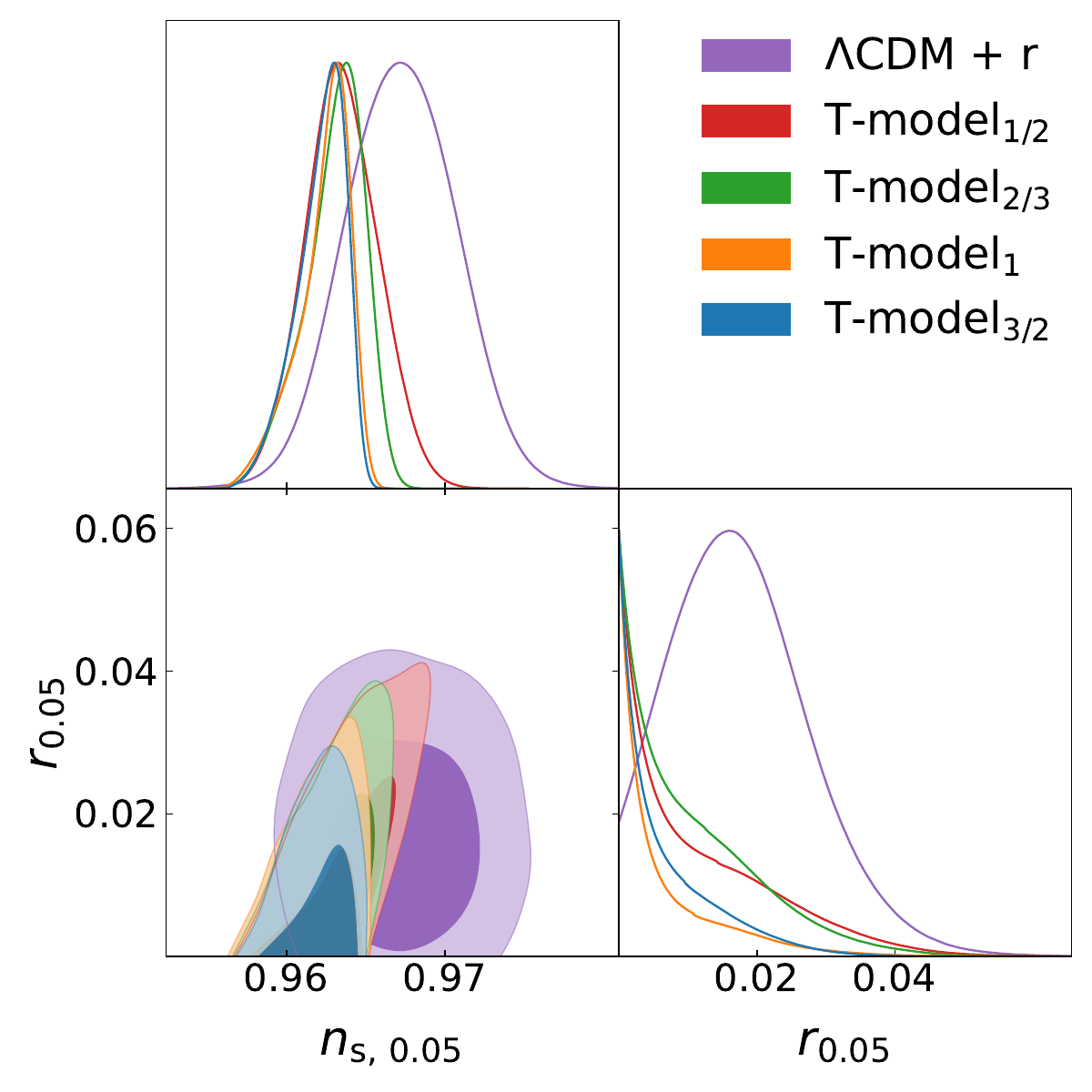}
\includegraphics[width=0.49\textwidth]{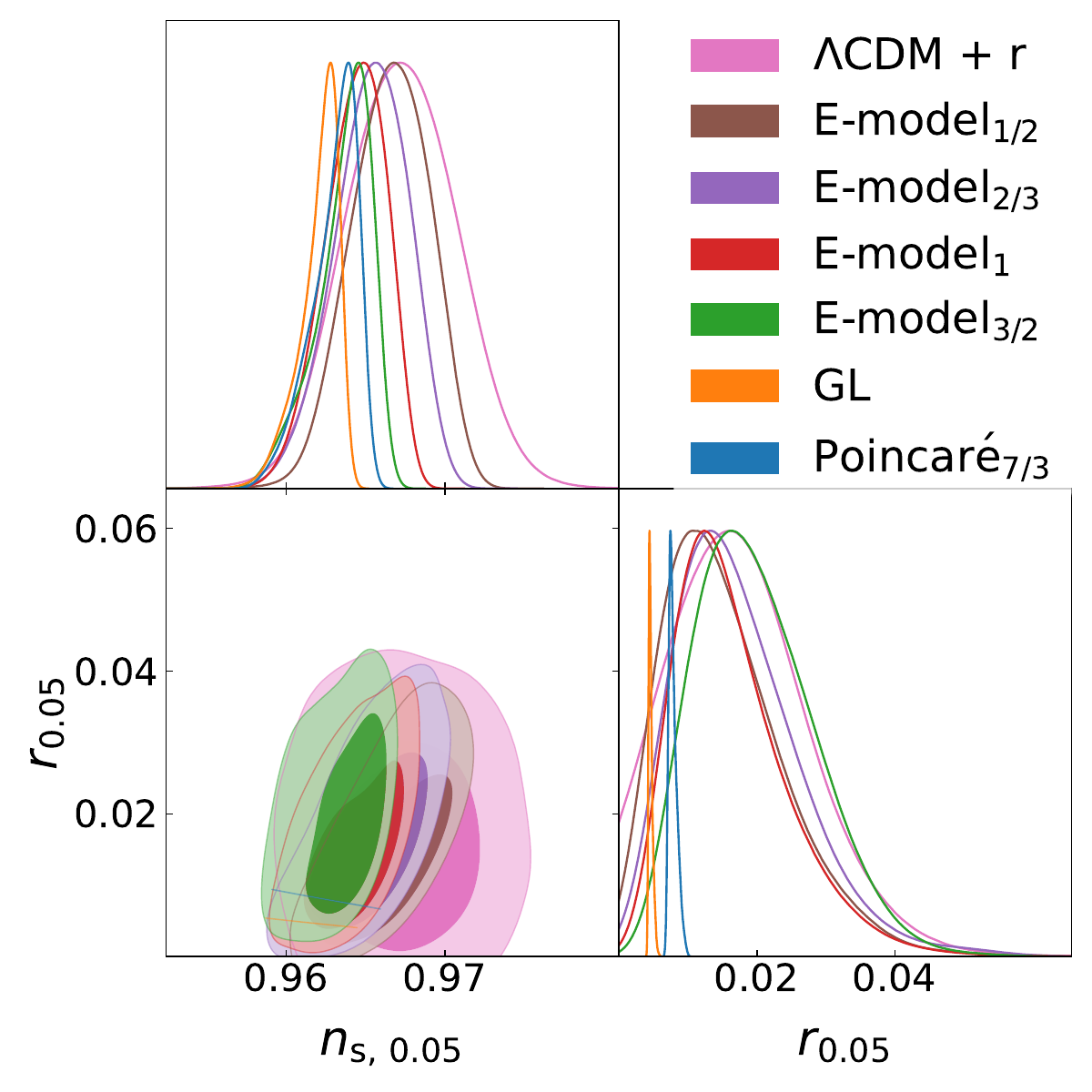}
\includegraphics[width=0.49\textwidth]{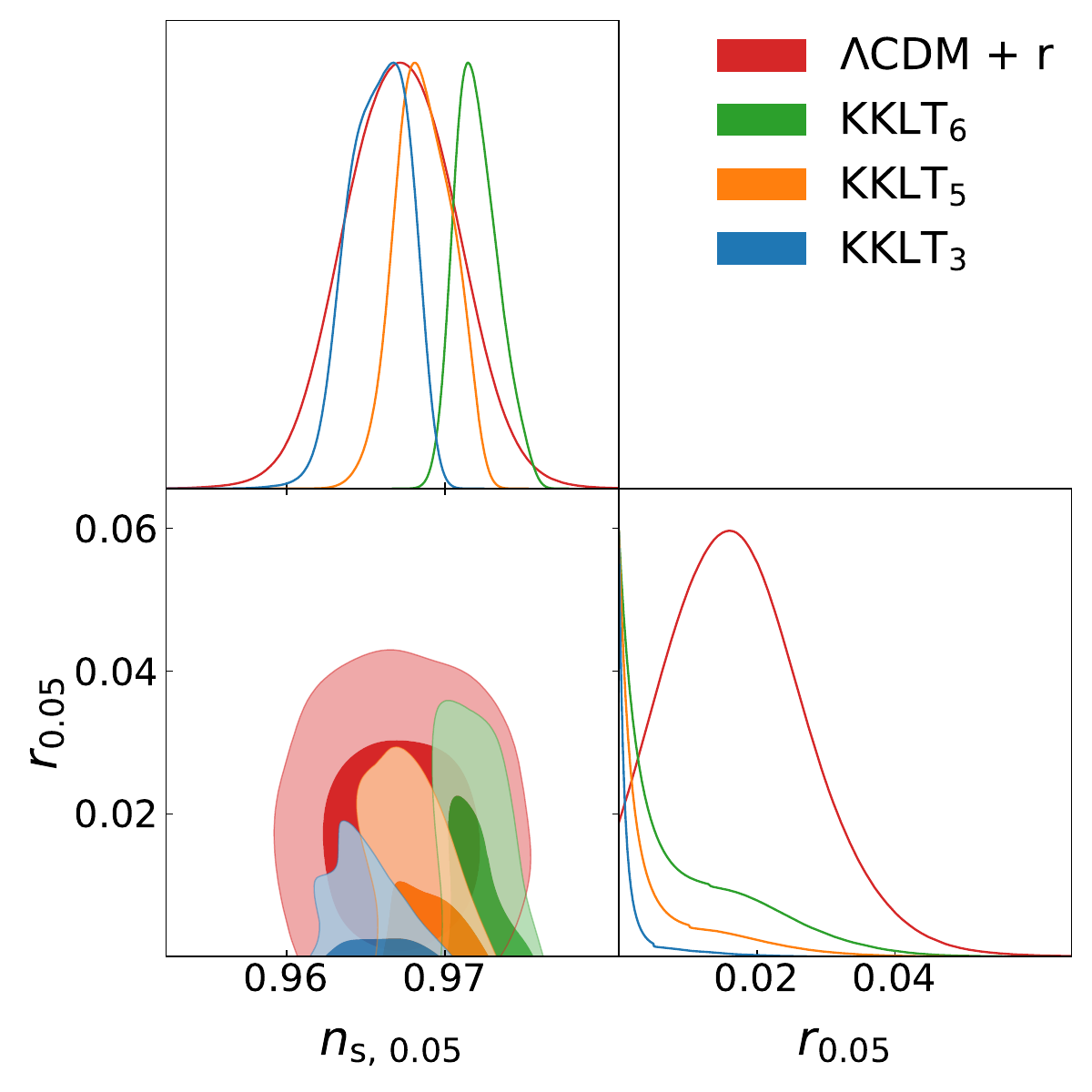}
\caption{Marginalised joint confidence contours for the scalar spectral index $n_{\rm s,\,0.05}$ and 
tensor-to-scalar ratio $r_{0.05}$ for $R+R^2$ inflation (upper left), T-model of $\alpha$-attractor inflation 
(upper right), E-model of $\alpha$-attractor inflation (lower left), and KKLT inflation (lower right), 
at 68\% CL and 95\% CL. In the E-model panel we include the contours for GL inflation corresponding to 
$n = 1$ and $\alpha = 1/9$ \cite{Goncharov:1983mw,Linde:2014hfa} and for Poincar\'e disk inflation with $n=1$ 
and $\alpha^{\rm E} = 7/3$ \cite{Ferrara:2016fwe,Kallosh:2019hzo}.
Here reheating parameters correspond to $w_{\rm re} = 0$ and $\rho_{\rm re}^{1/4} > 1\,{\rm TeV}$.
The star in the upper left panel corresponds to the standard prediction in $R+R^2$ inflation assuming 
the values for reheating  from \cite{Gorbunov:2010bn,Bezrukov:2011gp}.}
\label{fig:nsr_w0}
\end{figure}
In Fig.~\ref{fig:nsr_w0}, we show the 68\% CL and 95\% CL posterior distributions of the scalar spectral index 
$n_{\rm s,\,0.05}$ and tensor-to-scalar ratio $r_{0.05}$ for $R+R^2$ inflation, T-model and E-model of $\alpha$-attractor 
inflation for $n = 1/2,\,2/3,\,1,\,3/2$, and KKLT inflation for $p = 3,\,5,\,6$, for the baseline reheating scenario. 
We collect constraints and mean values on the inflationary parameters in Table~\ref{tab:results_w0}.
\begin{table*}
\centering
\begin{tabular}{l|cccc}
\hline
\hline
Parameter                                        & $\Lambda$CDM+$r$              & $R+R^2$          & GL    & Poincar\'e$_{7/3}$ \\
\hline
$\ln \left(  10^{10} A_{\rm s} \right)$          & $3.048^{+0.012}_{-0.014}$ & $3.048\pm 0.012$         & $3.046\pm 0.013$  & $3.049\pm 0.014$ \\
$\ln \left( \rho_{\rm re}/M_{\rm Pl}^4 \right)$ (at 95\% CL) & $-$                       & $> -118$     & $> -91.5$ & $> -106$ \\
\hline
$n_{\rm s,\,0.05}$                                & $0.9672\pm 0.0035$        & $0.9634^{+0.0021}_{-0.0011}$ & $0.9621^{+0.0015}_{-0.0007}$ & $0.9630^{+0.0018}_{-0.0009}$ \\
$r_{0.05}$                            (at 95\% CL)             & $< 0.036$    & $0.0038^{+0.0002}_{-0.0004}$ & $0.0046^{+0.0002}_{-0.0003}$ & $0.0078^{+0.0013}_{-0.0009}$ \\
$N_{0.05}$                                        & $-$                       & $52.0^{+3.0}_{-1.6}$ & $53.1^{+2.1}_{-1.1}$ & $52.7^{+2.6}_{-1.4}$ \\
$\log \left(T_{\rm re}/{\rm GeV}\right)$  (at 95\% CL)                       & $-$                       & $> 5.0$ & $> 7.8$    & $> 6.2$ \\
\hline
\end{tabular}
\newline
\vspace*{.2 cm}
\newline
{\small
\begin{tabular}{l|cccc}
\hline
\hline
Parameter                                        & T-model$_{1/2}$           & T-model$_{2/3}$     & T-model$_{1}$     & T-model$_{3/2}$ \\
\hline
$\ln \left(  10^{10} A_{\rm s} \right)$          & $3.047^{+0.012}_{-0.013}$ & $3.051\pm 0.013$   & $3.047\pm 0.014$     & $3.049\pm 0.013$ \\
$\ln \left( \rho_{\rm re}/M_{\rm Pl}^4 \right)$ (at 95\% CL) & $> -118$      & $> -113$  & $> -113$ & $> -108$ \\
$\alpha^{\rm T}$ (at 95\% CL)                           & $< 14.7$     & $< 10.6$ & $< 7.3$  & $< 6.3$ \\
\hline
$n_{\rm s,\,0.05}$                                & $0.9635\pm 0.0023$ & $0.9630^{+0.0023}_{-0.0014}$ & $0.9624^{+0.0020}_{-0.0010}$    & $0.9622^{+0.0019}_{-0.0011}$ \\
$r_{0.05}$    (at 95\% CL)                                    & $< 0.034$      & $< 0.030$  & $< 0.024$   & $< 0.023$ \\
$N_{0.05}$                                        & $51.9^{+2.9}_{-1.9}$         & $52.3^{+2.8}_{-1.6}$ & $52.2^{+2.8}_{-1.5}$  & $52.3^{+2.7}_{-1.7}$ \\
$\log \left(T_{\rm re}/{\rm GeV}\right)$     (at 95\% CL)   & $> 5.0$                   & $> 5.5$            & $> 5.5$        & $> 6.0$  \\
\hline
\end{tabular}}
\newline
\vspace*{.2 cm}
\newline
{\small
\begin{tabular}{l|cccc}
\hline
\hline
Parameter                                        & E-model$_{1/2}$           & E-model$_{2/3}$     & E-model$_{1}$     & E-model$_{3/2}$ \\
\hline
$\ln \left(  10^{10} A_{\rm s} \right)$          & $3.051\pm 0.013$ & $3.050\pm 0.014$   & $3.049\pm 0.014$     & $3.049\pm 0.013$ \\
$\ln \left( \rho_{\rm re}/M_{\rm Pl}^4 \right)$ (at 95\% CL)  & $-$     & $-$ & $> -108$  & $> -108$  \\
$\alpha^{\rm E}$  (at 95\% CL)                          & $< 30.9$     & $< 25.9$  & $< 16.8$ & $< 16.1$ \\
\hline
$n_{\rm s,\,0.05}$                                & $0.9666^{+0.0027}_{-0.0023}$ & $0.9654^{+0.0026}_{-0.0021}$ & $0.9643^{+0.0022}_{-0.0017}$    & $0.9637^{+0.0021}_{-0.0011}$ \\
$r_{0.05}$                      (at 95\% CL)                  & $< 0.032$     & $0.017^{+0.018}_{-0.016}$  & $0.016^{+0.017}_{-0.014}$  & $0.020^{+0.017}_{-0.015}$ \\
$N_{0.05}$                                        & $51.8^{+3.2}_{-2.0}$         & $51.6^{+3.0}_{-2.2}$ & $52.7^{+2.7}_{-1.6}$  & $53.3^{+2.6}_{-1.3}$ \\
$\log \left(T_{\rm re}/{\rm GeV}\right)$  (at 95\% CL)      & $-$                    & $-$          & $> 5.8$        & $> 6.1$  \\
\hline
\end{tabular}}
\newline
\vspace*{.2 cm}
\newline
\begin{tabular}{l|ccc}
\hline
\hline
Parameter                                        & KKLT$_3$                & KKLT$_5$ & KKLT$_6$\\
\hline
$\ln \left(  10^{10} A_{\rm s} \right)$          & $3.057^{+0.012}_{-0.014}$    & $3.056\pm 0.013$   & $3.055^{+0.015}_{-0.017}$ \\
$\ln \left( \rho_{\rm re}/M_{\rm Pl}^4 \right)$ (at 95\% CL)  & $-$            & $-$             & $< -42.6$  \\
$m\,[{\rm M_{\rm Pl}}]$  (at 95\% CL)                                     & $< 5.6$         & $< 5.1$  & $< 6.8$  \\
\hline
$n_{\rm s,\,0.05}$                                & $0.9659^{+0.0021}_{-0.0017}$ & $0.9687\pm 0.0018$ & $0.9720^{+0.0011}_{-0.0016}$ \\
$r_{0.05}$ (at 95\% CL)                                       & $< 0.011$       & $< 0.023$  & $< 0.030$  \\
$N_{0.05}$                                        & $49.1^{+2.6}_{-3.1}$         & $48.9^{+2.4}_{-3.4}$ & $48.6^{+1.7}_{-3.2}$ \\
$\log \left(T_{\rm re}/{\rm GeV}\right)$ (at 95\% CL)       & $-$                   & $-$           & $< 13.1$  \\
\hline
\end{tabular}
\caption{\label{tab:results_w0} 
Constraints on the main and derived parameters (at 68\% CL if not otherwise stated) for $\Lambda$CDM+$r$, $R+R^2$, GL, Poincar\'e, 
$\alpha$-attractor inflation, and KKLT inflation considering the combination P18+BK18+BAO+RSD+SNe.}
\end{table*}

$R+R^2$ marginalised posterior distributions are well in agreements with the reference posteriors obtained for the 
phenomenological power-law case with a predicted value of the scalar spectral index slightly lower than the 68\% CL; 
see Fig.~\ref{fig:nsr_w0}.
The predictions in the basic version of the $R + R^2$ (corresponding to the star in Fig.~\ref{fig:nsr_w0}) gives a smaller 
value of the scalar spectral index compared to $\alpha$-attractors.
The reason is that in the basic $R + R^2$ model the only interactions are gravitational, and therefore reheating is 
inefficient, which leads to a smaller number of $e$-folds and consequently a smaller scalar spectral index. For comparison,
reheating in the Higgs inflation is very efficient, leading to a larger value of the scalar spectral index \cite{Bezrukov:2011gp}.

$\alpha$-attractor inflation and KKLT inflation cover a larger portion of the $n_{\rm s}$-$r$ parameter space thank to 
the extra parameter. While the latter covers better the left part of the contour plot, the former covers larger values 
of the scalar spectral index.
E-model $\alpha$-attractor inflation is able to fit the hint of non-zero primordial gravitational waves in BK18 
observations with a value of the scalar spectral index compatible to the other cosmological datasets (mostly driven by P18 data).

While we consider $\alpha$ as a continuous parameter in our analysis, there are examples, such as in advanced supergravity models, 
where $\alpha$ assumes discrete values; see Refs.~\cite{Ferrara:2016fwe,Kallosh:2019hzo}. 
We consider a specific case of Poincar\'e disk inflation with $n=1$ and $\alpha^{\rm E} = 7/3$, 
with zero extra parameters as $R+R^2$ and GL inflation; see Fig.~\ref{fig:nsr_w0} and Table~\ref{tab:results_w0}.

The reheating energy density parameter $\ln (\rho_{\rm re}/M_{\rm Pl}^4)$ is often unconstrained since all the inflationary 
models considered here are well in agreement with cosmological observations as can be seen in Figs.~\ref{fig:reheating_R2} and 
\ref{fig:reheating_models} (dotted lines). Indeed, there is no need of a specific reheating 
behaviour to accommodate their predictions.
However, there is a preference for a short reheating period for $R+R^2$ and $\alpha$-attractor inflation while a 
longer period seems preferred for KKLT inflation; see Figs.~\ref{fig:reheating_R2} and \ref{fig:reheating_models}.

For $\alpha$-attractor and KKLT inflation the constraints on the additional inflationary parameter correspond roughly to 
$\alpha^{\rm T} \lesssim 20$, $\alpha^{\rm E} \lesssim 30$, and $m/M_{\rm Pl} \lesssim 10$ at 95\% CL, 
see Table~\ref{tab:results_w0}; see Ref.~\cite{Iacconi:2023mnw} for a detailed study on the impact of linear and logarithmic prior 
on $\alpha^{\rm T}$.

\begin{figure}
\centering
\includegraphics[width=0.49\textwidth]{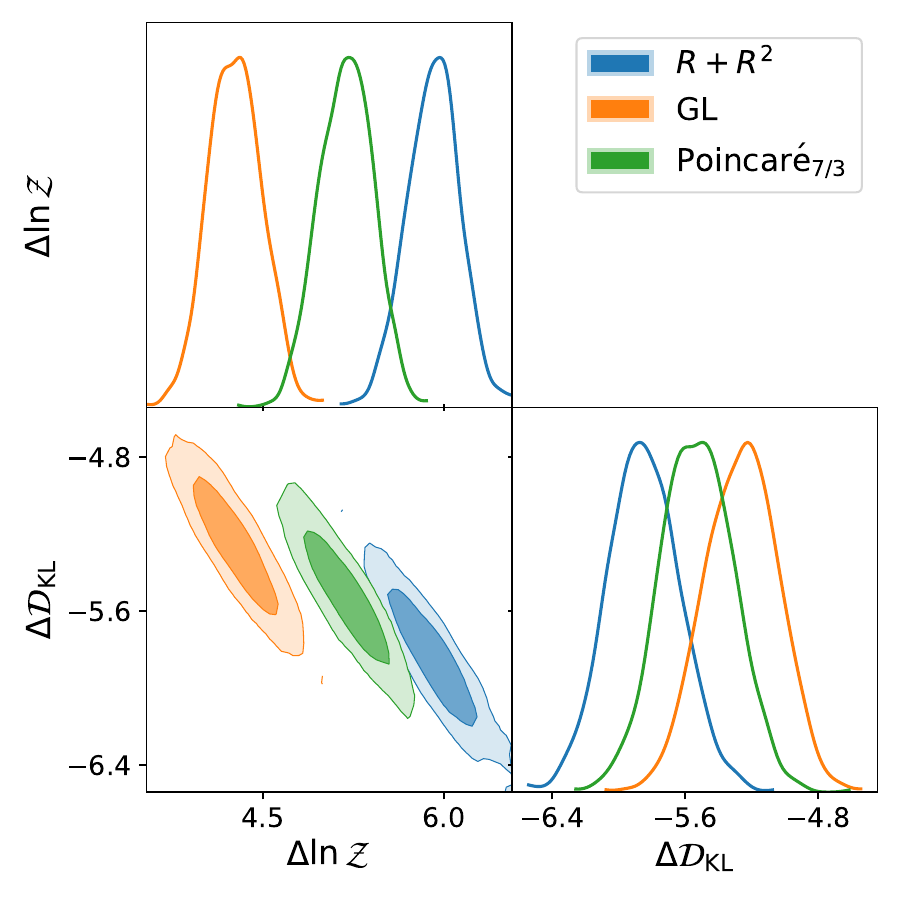}
\includegraphics[width=0.49\textwidth]{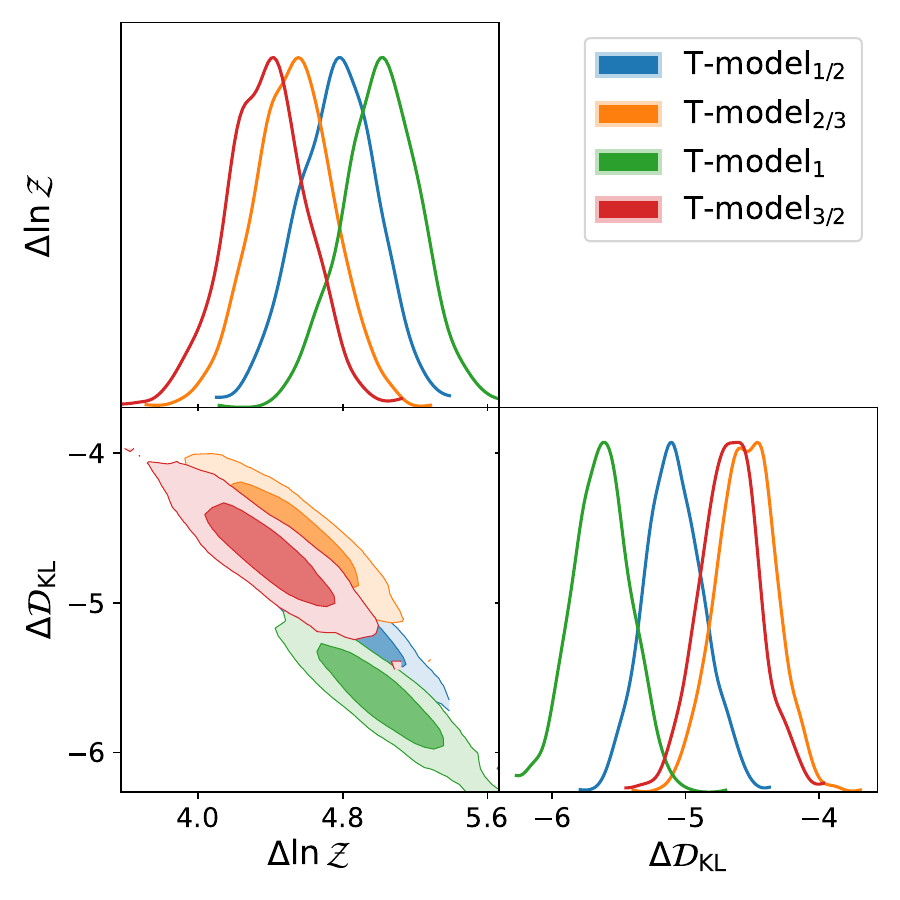} 
\\
\includegraphics[width=0.49\textwidth]{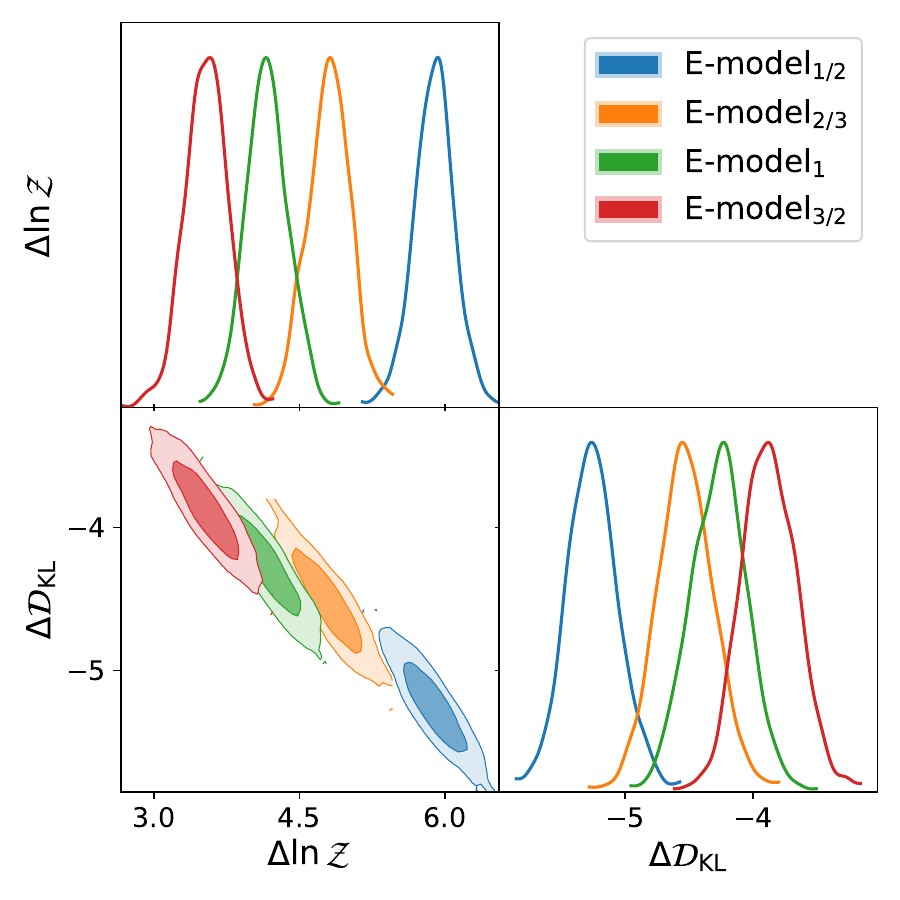}
\includegraphics[width=0.49\textwidth]{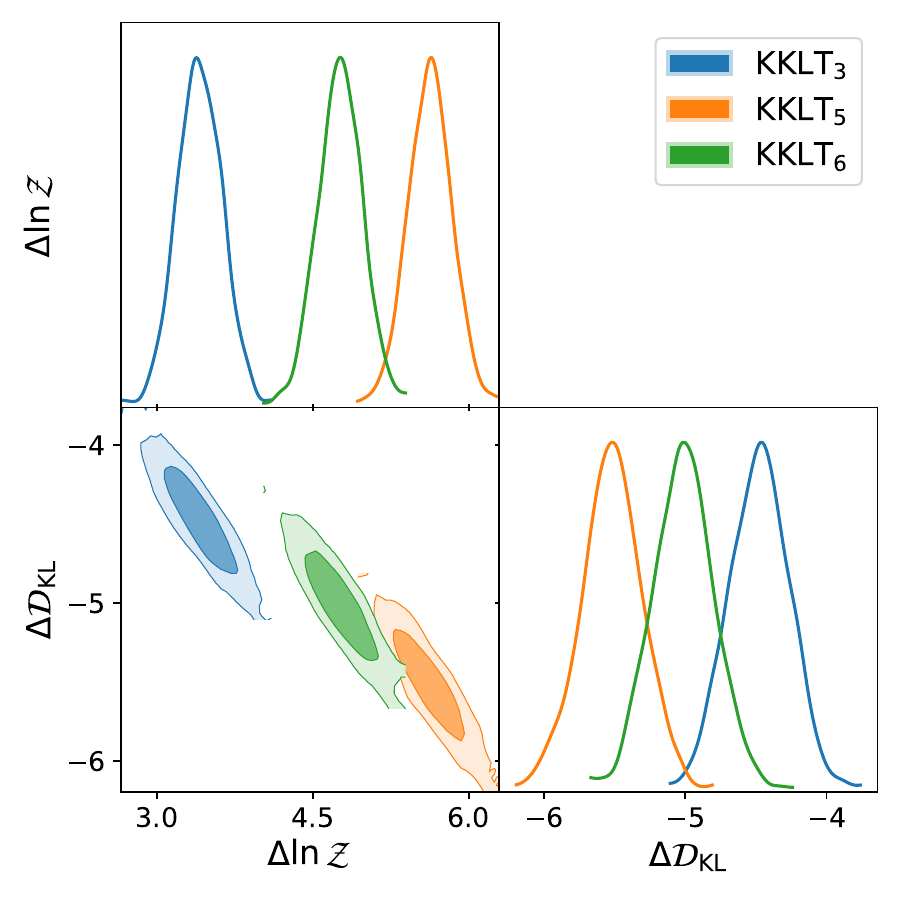}
\caption{Relative log-evidence $\Delta \ln {\cal Z}$ and relative KL divergence $\Delta {\cal D}_{\rm KL}$ for the 
$R+R^2$, GL, Poincar\'e disk, $\alpha$-attractor, and KKLT inflation with respect to $\Lambda$CDM+$r$ assuming baseline reheating scenario.
\label{fig:stats}}
\end{figure}

We investigate the log-evidence and KL divergence for the inflationary model analysed in comparison 
to the $\Lambda$CDM+$r$ model \cite{Hergt:2018ksk,Hergt:2022fxk}. We calculate the evidence 
${\cal Z} \equiv P({\cal D}|{\cal M})$, that is the marginal likelihood for the model ${\cal M}$, and we report the Bayes' 
factors as 
\begin{equation}
    \Delta \ln {\cal Z} = \ln \frac{P({\cal D}|{\cal M})}{P({\cal D}|{\cal M}_{\Lambda {\rm CDM}+r})} \,,    
\end{equation}
where the evidence is given by 
\begin{equation}
    {\cal Z} \equiv P({\cal D}|{\cal M}) = \int \dd \theta P({\cal D}|\theta,{\cal M}) \pi(\theta) \,.
\end{equation}
and $\Delta \log {\cal Z} > 0$ favours the reference model, here the $\Lambda$CDM+$r$. 
$P({\cal D}|\theta,{\cal M})$ is the likelihood of the parameters $\theta$ given the data ${\cal D}$. 
In order to quantify going from the prior distribution to the 
posterior distribution, we calculated the relative entropy or KL divergence
\begin{equation}
    {\cal D}_{\rm KL} = \int \dd \theta P(\theta|{\cal D},{\cal M}) \ln \left(\frac{P(\theta|{\cal D},{\cal M})}{\pi(\theta)}\right)
\end{equation}
where $P(\theta|{\cal D},{\cal M})$ is the posterior distribution of the parameters $\theta$ and 
$\pi(\theta)$ are the prior ranges on the parameters.
We show these quantities in Fig.~\ref{fig:stats} in a triangle plot, for all the inflationary models analysed.
See Refs.~\cite{Hergt:2018ksk,Hergt:2022fxk,Martin:2024qnn} for an extended discussion on KL divergence and 
other Bayesian estimators in the context of inflationary models.

Looking at the relative log-evidence, we can see a clear preference for the inflationary models studied 
compared to the $\Lambda$CDM+$r$ model. The inflationary models that perform better for each class are $R+R^2$ inflation 
with $\Delta \ln {\cal Z} = 5.9 \pm 0.2$, T-model $\alpha$-attractor inflation for $n=1$ with $\Delta \ln {\cal Z} = 5.0 \pm 0.2$, 
E-model $\alpha$-attractor inflation for $n=1/2$ with $\Delta \ln {\cal Z} = 5.9 \pm 0.2$, and KKLT inflation for 
$p=5$ with $\Delta \ln {\cal Z} = 5.6 \pm 0.2$. 
Comparing the different models no one results preferred according to the revised Jeffrey's scale \cite{Kass:1995loi};  
we always have for any pair of inflationary models $\Delta \ln {\cal Z} < 2.5$. 
This estimator tends to penalise the addition of parameters which usually leads to spread the 
model's predictive probability over a larger parameter space. Here, $R+R^2$, GL, and Poincar\'e disk inflation have six cosmological 
parameters, $\alpha$-attractor and KKLT inflation seven parameters, and the reference $\Lambda$CDM+$r$ seven as well. 
Note that, the inclusion of the tensor-to-scalar ratio, with prior range of $r_{0.05} \in [0,\,1]$, penalises the reference model 
compared to the standard $\Lambda$CDM model without primordial tensors which on the other hand would be preferred as discussed in 
Ref.~\cite{Hergt:2018ksk,Hergt:2021qlh}. Moreover, the scalar spectral index and the tensor-to-scalar ratio are derived parameters 
for the inflationary models in our analysis leading to an unavoidable use of 
different prior which might affect the model comparison, see Refs.~\cite{Hergt:2018ksk,Hergt:2022fxk}.

In terms of relative KL divergence, we can see no prior-to-posterior distribution compression for the inflationary 
model parameters. Indeed, while in $\Lambda$CDM+$r$ all the cosmological parameters are well constrained for the given prior ranges 
with a tight upper bound for the tensor-to-scalar ratio, for the inflationary models the reheating parameters 
are often unconstrained by current cosmological data.

\subsection{Effect of different reheating scenarios}
In this section, we compare the baseline reheating scenario studied in the previous subsection, in which the reheating phase 
last down to $\rho^{1/4}_{\rm re} = 1\,{\rm TeV}$ maximum and with $w_{\rm re} = 0$, to the following two reheating scenarios:
\begin{description}
    \item[restrictive] $\rho_{\rm re}^{1/4} > 1\,{\rm TeV},\quad -1/3 < w_{\rm re} < 1/3$ \,,
    \item[permissive] $\rho_{\rm re}^{1/4} > 10\,{\rm MeV},\quad -1/3 < w_{\rm re} < 1$ \,.\footnote{Similar and additional choices have been considered in Refs.~\cite{Planck:2013jfk,Hergt:2022fxk}.}
\end{description}
Here the energy density is bounded by requiring the reheating phase to end before electroweak scale ($\sim 10^2\, {\rm GeV}$) 
in the restrictive case and before Big Bang nucleosynthesis happens ($\sim 1\, {\rm MeV}$) \cite{deSalas:2015glj} in the 
permissive case.

Different reheating scenarios affect the inflationary predictions leading to different number of $e$-folds between horizon 
crossing and the end of inflation according to Eq.~\eqref{eqn:Nk}. 
The baseline reheating scenario studied in the previous subsection with 
$\rho_{\rm re}^{1/4} > 1\,{\rm TeV}$ and $w_{\rm re} = 0$ corresponds 
to $\Delta N \simeq -11.8$ maximum with respect to assume instantaneous reheating. Indeed, in the baseline reheating scenario 
we can have less $e$-folds compared to the assumption of instantaneous reheating.
For $w_{\rm reh}=0$, assuming a larger prior for the reheating energy density, as in the permissive case, 
corresponds to additional $\Delta N \simeq -3.8$ $e$-folds going from $1\,{\rm TeV}$ to $10\,{\rm MeV}$.
For $w_{\rm reh} \ne 0$ the situation is a bit more convolved. 
In the restrictive case, the prefactor $(1 - 3w_{\rm re})/(12 + 12w_{\rm re})$, which multiply $N_{\rm re}$, can vary in the 
range $[0,\,0.25]$. For $-1/3 < w_{\rm re} < 0$, we can have even less $e$-folds compared to the baseline case while 
keeping fix the reheating energy density. For $0 < w_{\rm re} < 1/3$ the impact of the reheating uncertainties on the 
inflationary predictions is reduced compared to the baseline case. In the permissive case, for $1/3 < w_{\rm re} < 1$, the 
prefactor changes sign leading in this case to a larger number of $e$-folds compared to instantaneous reheating.

\begin{figure}
\centering
\includegraphics[width=0.32\textwidth]{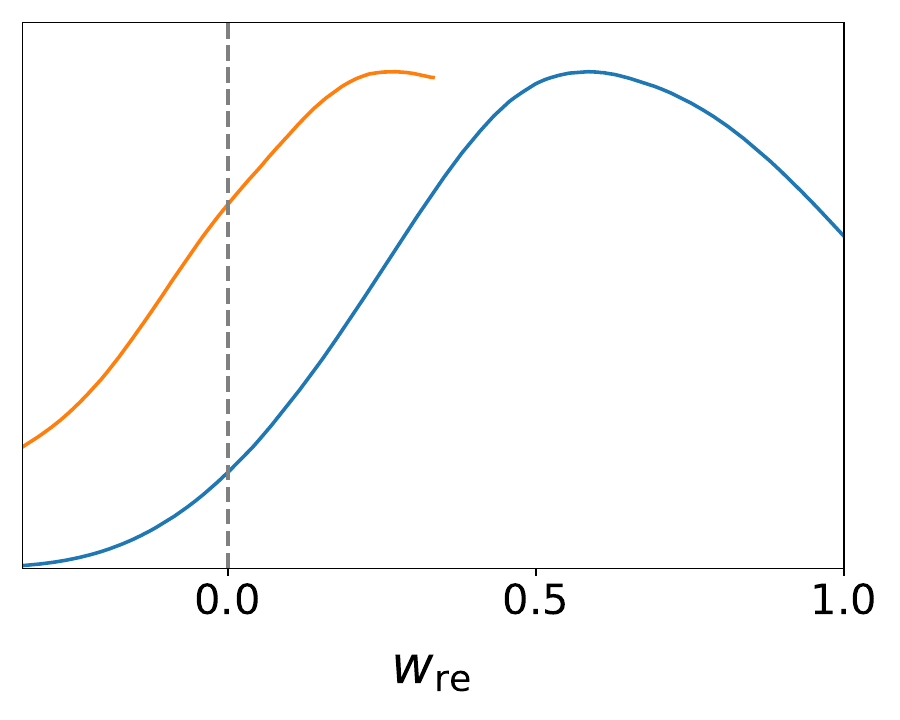}
\includegraphics[width=0.32\textwidth]{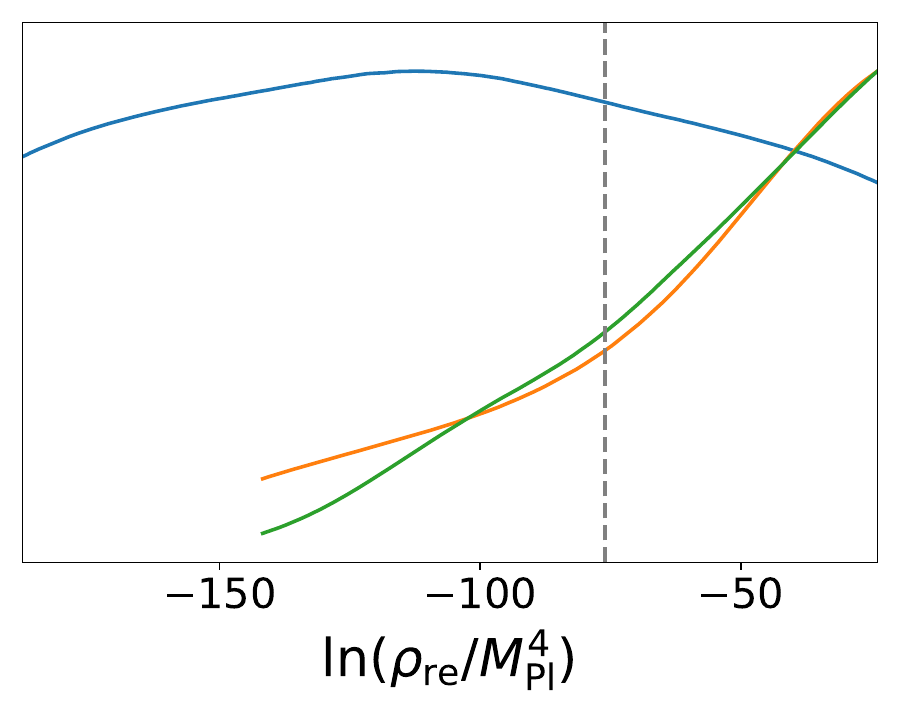}
\includegraphics[width=0.32\textwidth]{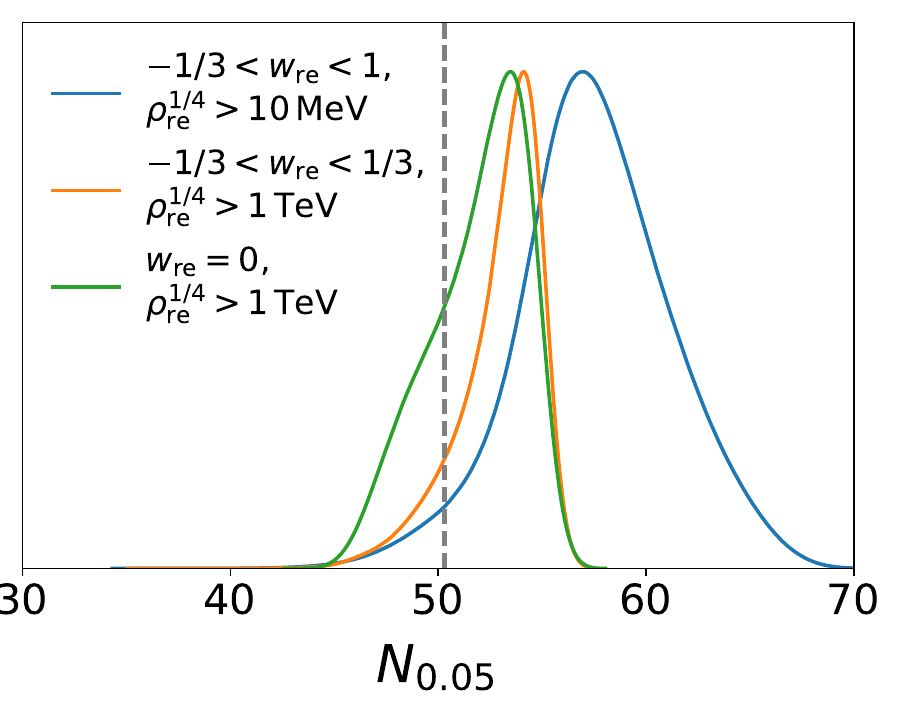}
\caption{Posterior distribution in $R+R^2$ inflation for the effective equation of state parameter $w_{\rm re}$ 
(left panel), the energy density at the end of reheating $\ln (\rho_{\rm re}/M_{\rm Pl}^4)$ (central panel), and 
the number of $e$-folds between the scale $k_*$ crosses the horizon and the end of inflation $N_{0.05}$ (right panel) 
for the minimal ($w_{\rm re} = 0,\, \rho^{1/4}_{\rm re} > 1\,{\rm TeV}$), restricted 
($-1/3 < w_{\rm re} < 1/3,\, \rho^{1/4}_{\rm re} > 1\,{\rm TeV}$), and permissive  
($-1/3 < w_{\rm re} < 1,\, \rho^{1/4}_{\rm re} > 10\,{\rm MeV}$) reheating scenario in blue, orange, and green, respectively. 
The dashed vertical lines correspond to the standard value for reheating in $R+R^2$ inflation from 
\cite{Gorbunov:2010bn,Bezrukov:2011gp}.}
\label{fig:reheating_R2}
\end{figure}

\begin{figure}
\centering
\includegraphics[width=0.32\textwidth]{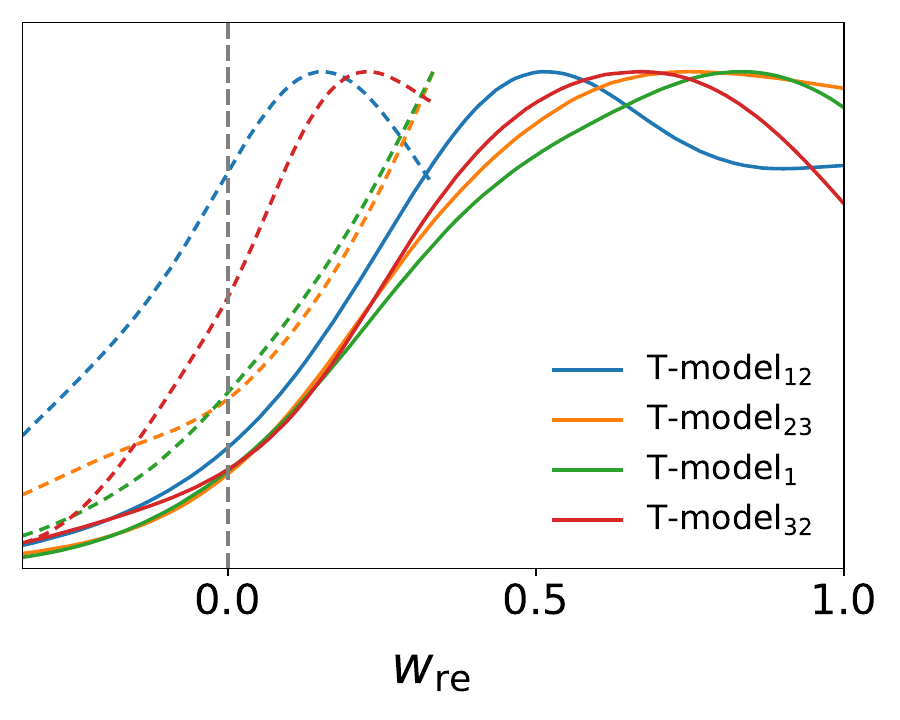}
\includegraphics[width=0.32\textwidth]{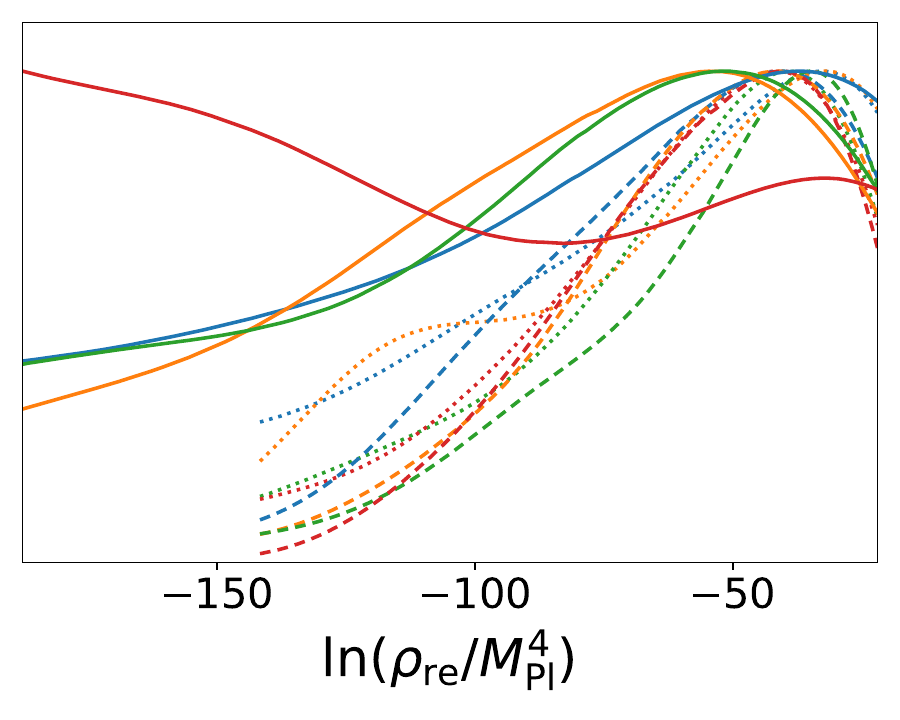}
\includegraphics[width=0.32\textwidth]{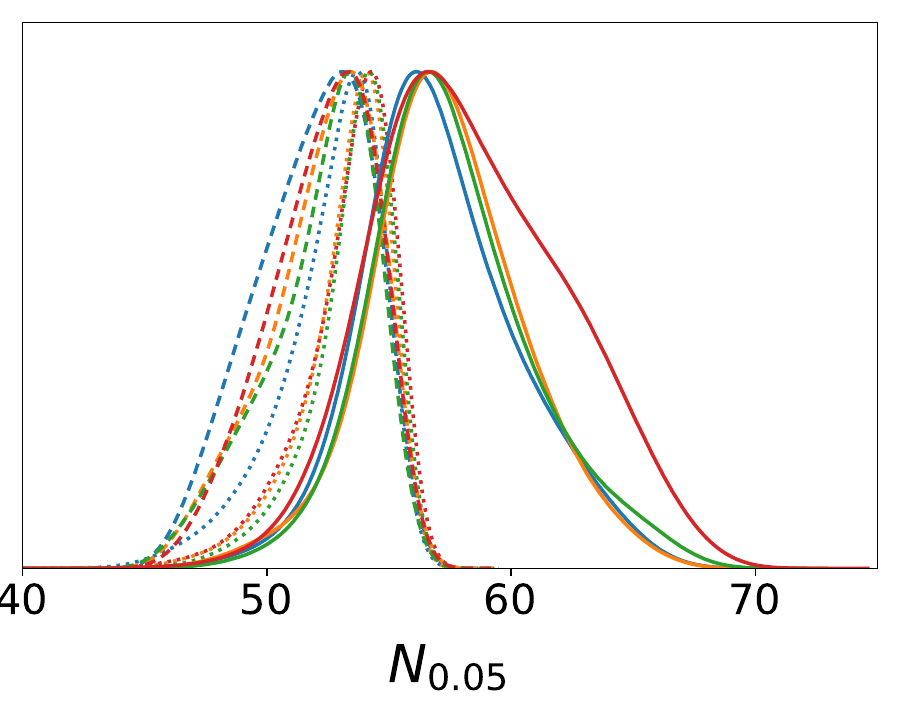}
\\
\includegraphics[width=0.32\textwidth]{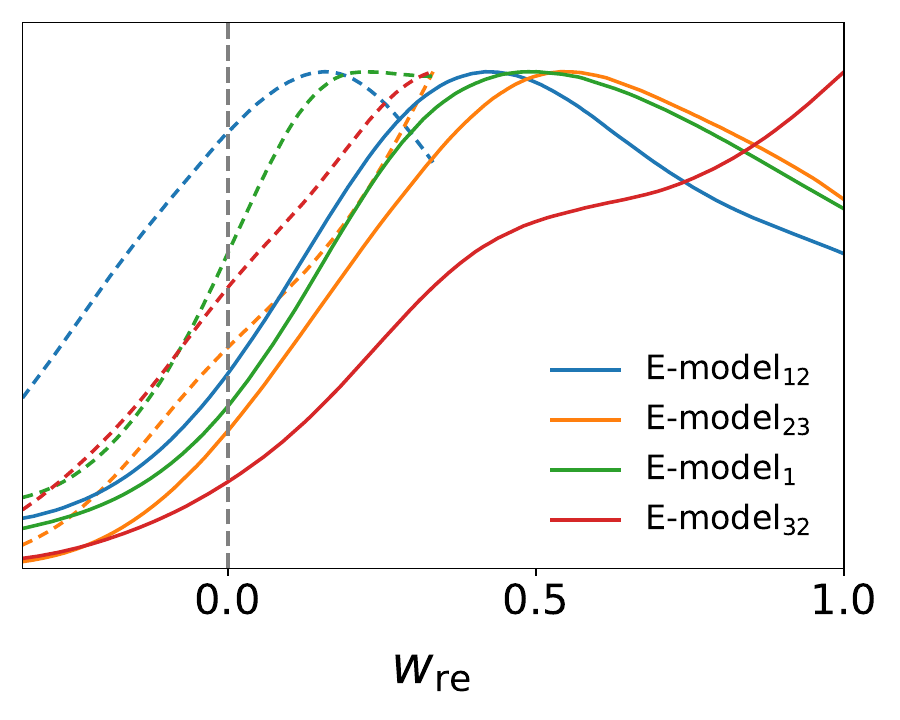}
\includegraphics[width=0.32\textwidth]{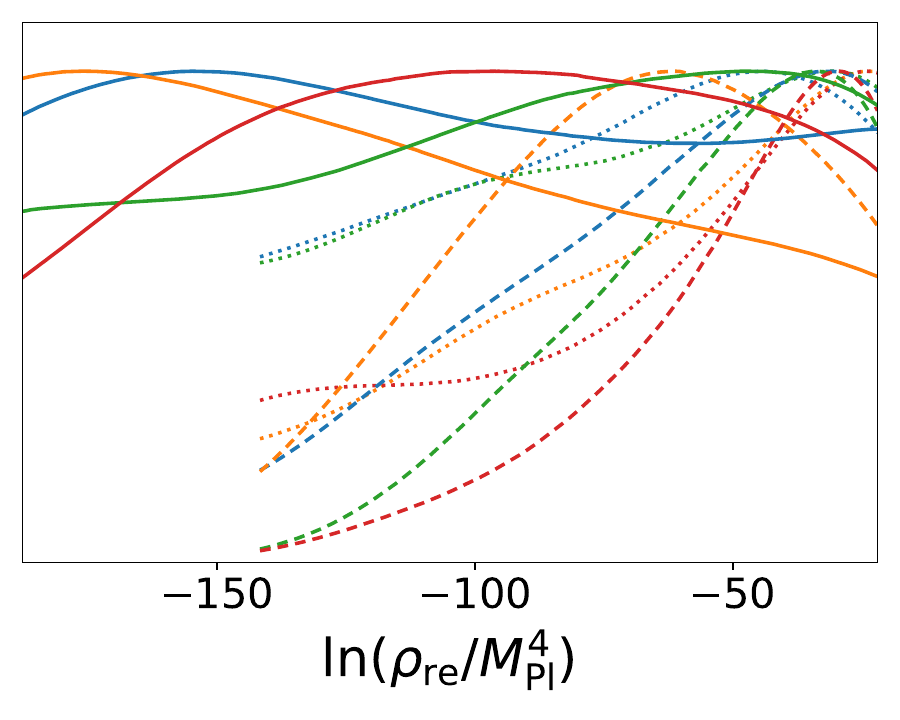}
\includegraphics[width=0.32\textwidth]{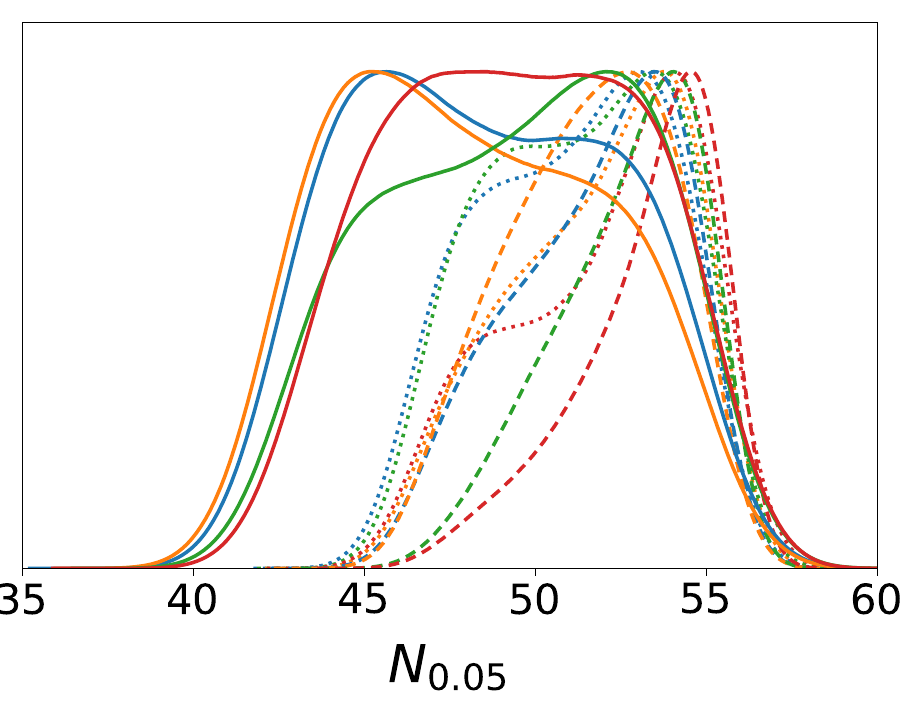}
\\
\includegraphics[width=0.32\textwidth]{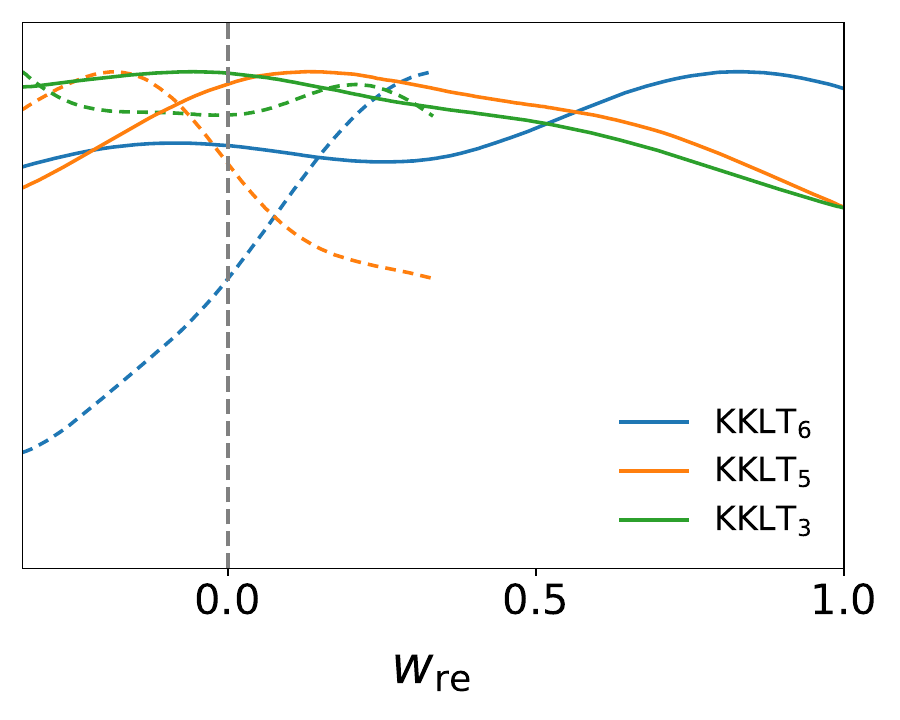}
\includegraphics[width=0.32\textwidth]{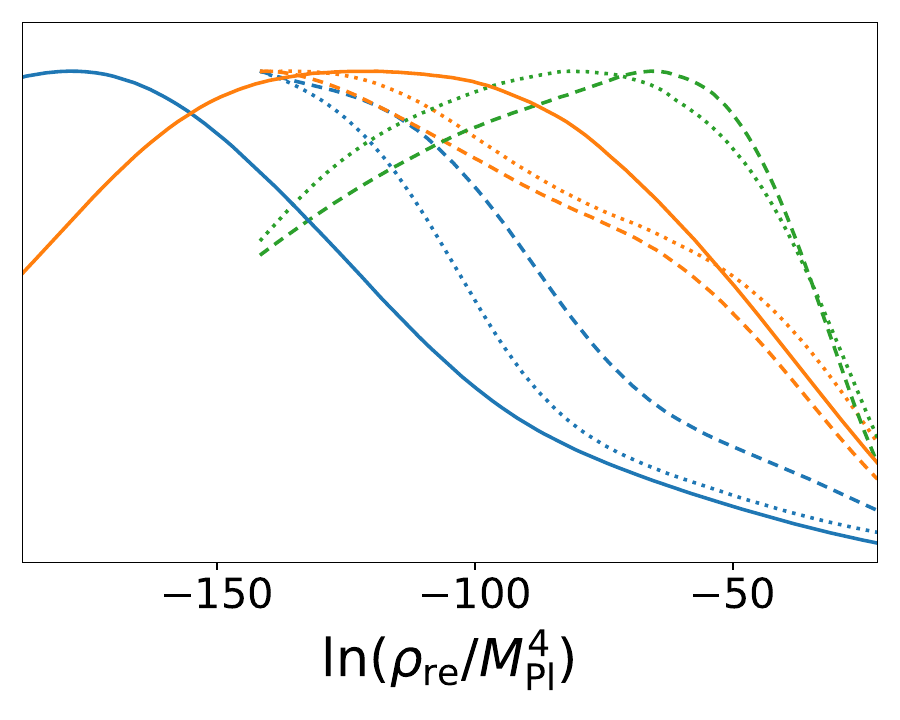}
\includegraphics[width=0.32\textwidth]{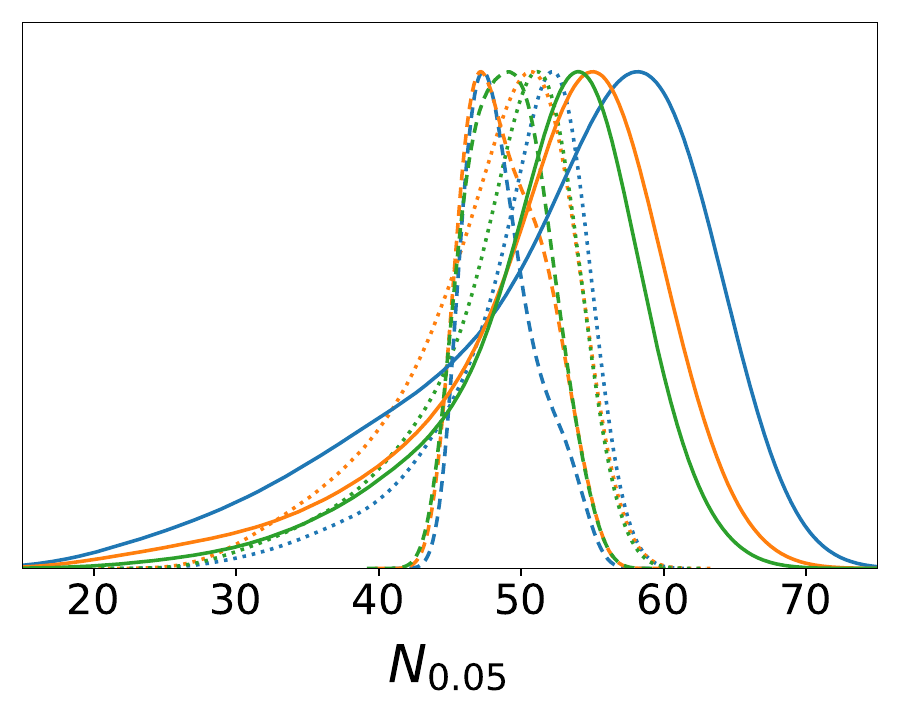}
\caption{Same as Fig.~\ref{fig:reheating_R2} for T-model of $\alpha$-attractor inflation (upper row), 
E-model of $\alpha$-attractor inflation (central row), and KKLT inflation (lower row). Solid lines 
correspond to the permissive reheating scenario, dashed ones to the restricted reheating scenario, and 
the dotted ones to our baseline reheating scenario.}
\label{fig:reheating_models}
\end{figure}

In Figs.~\ref{fig:reheating_R2} and \ref{fig:reheating_models} the posterior distributions of the effective equation 
of state parameter $w_{\rm re}$ (for the restrictive and permissive reheating), the energy density at the end of reheating 
$\ln (\rho_{\rm re}/M_{\rm Pl}^4)$ (for the baseline, the restrictive, and the permissive reheating), and the number of $e$-folds 
$N_{0.05}$ between the scale $k_* = 0.05\,{\rm Mpc}^{-1}$ crosses the horizon and the end of inflation (for the baseline, 
the restrictive, and the permissive reheating) are shown; see Tables~\ref{tab:results_wrest} and \ref{tab:results_wperm} for 
means and uncertainties on the inflationary parameters assuming restrictive and permissive reheating scenarios, respectively.

\begin{figure}
\centering
\includegraphics[width=0.8\textwidth]{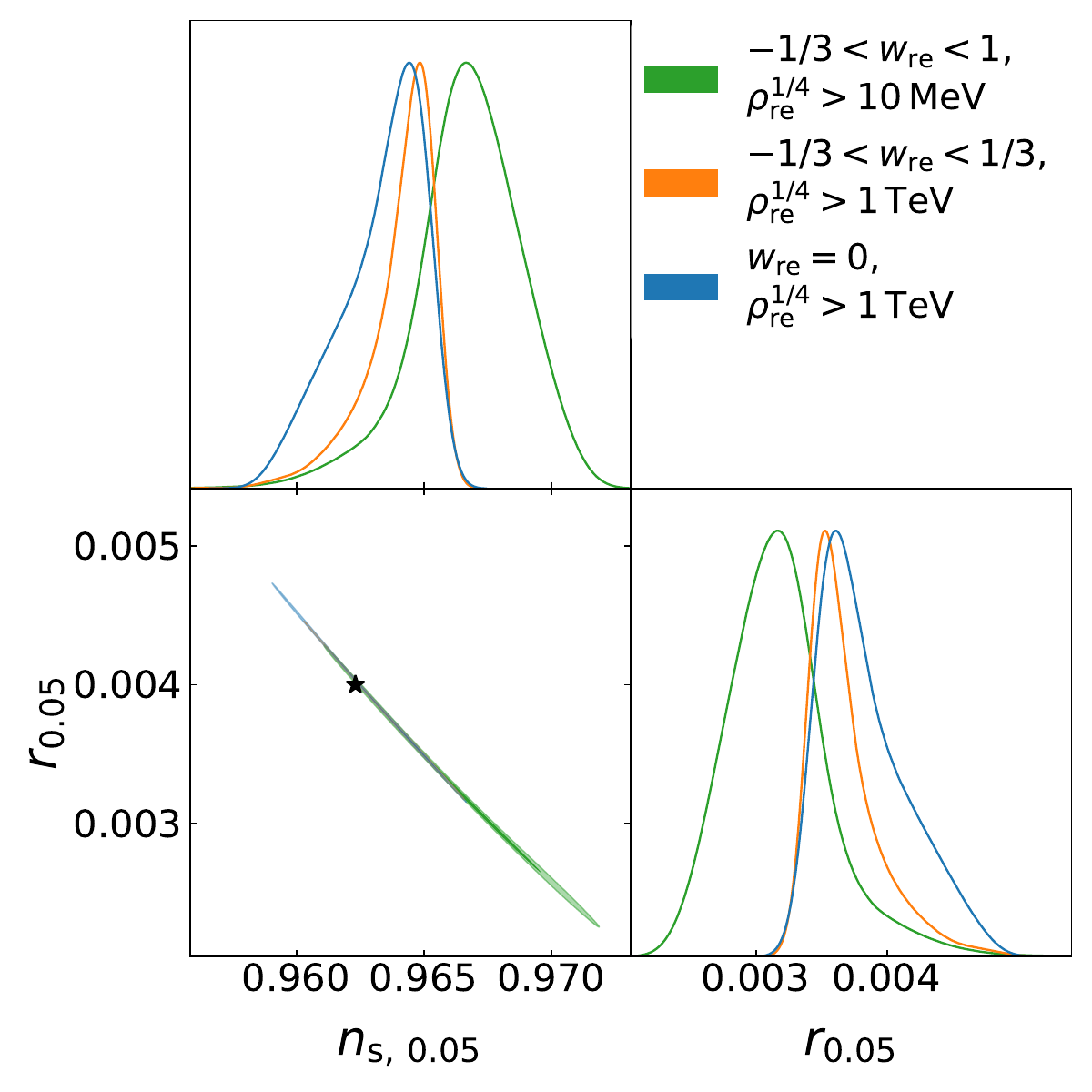}
\caption{Marginalised joint confidence contours for the scalar spectral index $n_{\rm s,\,0.05}$ and tensor-to-scalar ratio 
$r_{0.05}$ for $R+R^2$ inflation at 68\% CL and 95\% CL for the baseline ($w_{\rm re} = 0,\, \rho^{1/4}_{\rm re} > 1\,{\rm TeV}$), 
restricted ($-1/3 < w_{\rm re} < 1/3,\, \rho^{1/4}_{\rm re} > 1\,{\rm TeV}$), and permissive  
($-1/3 < w_{\rm re} < 1,\, \rho^{1/4}_{\rm re} > 10\,{\rm MeV}$) reheating scenario in blue, orange, and green, respectively. 
Here reheating parameters correspond to $w_{\rm re} = 0$ and $\rho_{\rm re}^{1/4} > 1\,{\rm TeV}$. The star corresponds to the 
prediction standard in $R+R^2$ inflation assuming standard values for reheating  from 
\cite{Gorbunov:2010bn,Bezrukov:2011gp}.\label{fig:nsr_wre_R2}}
\end{figure}

\begin{figure}
\centering
\includegraphics[width=0.49\textwidth]{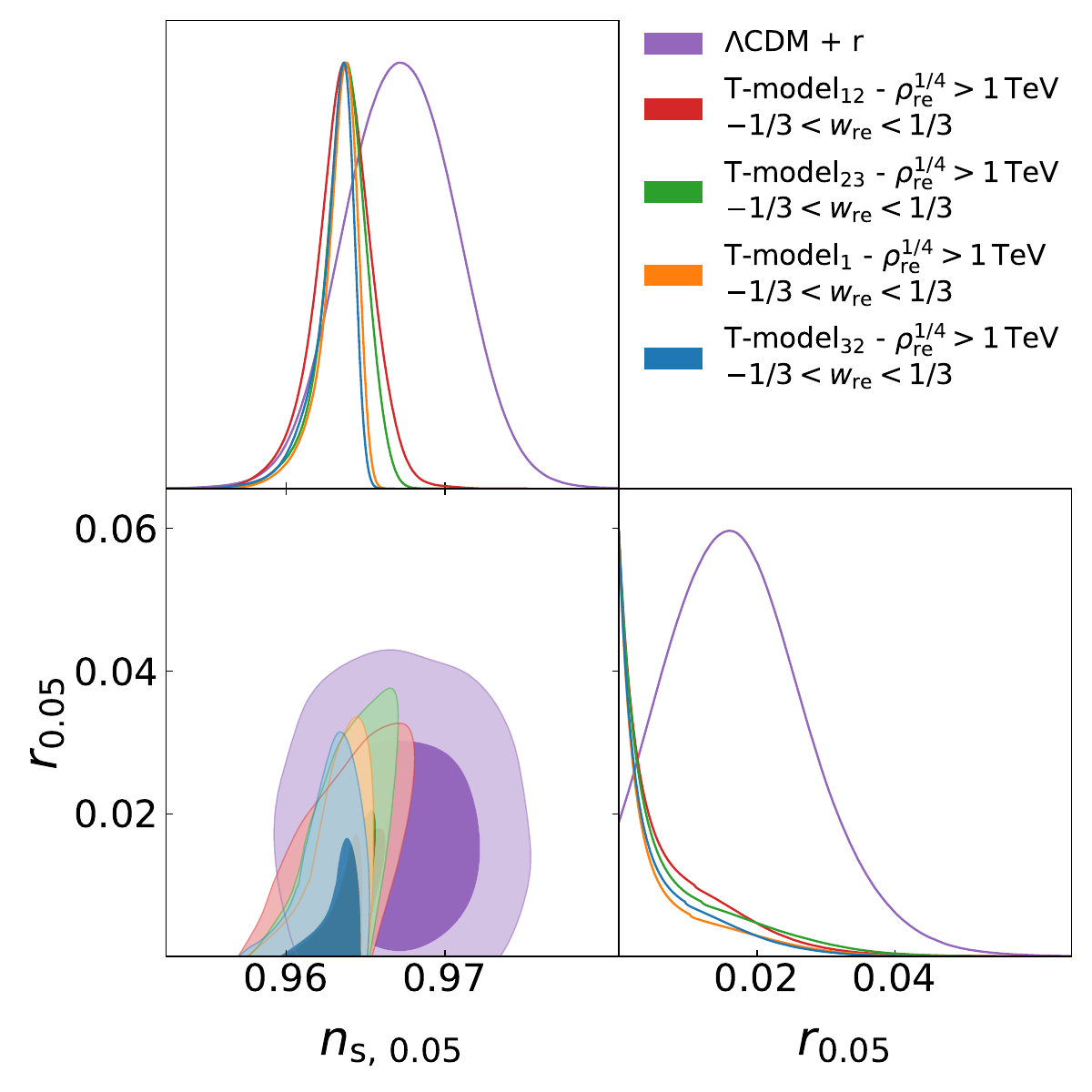}
\includegraphics[width=0.49\textwidth]{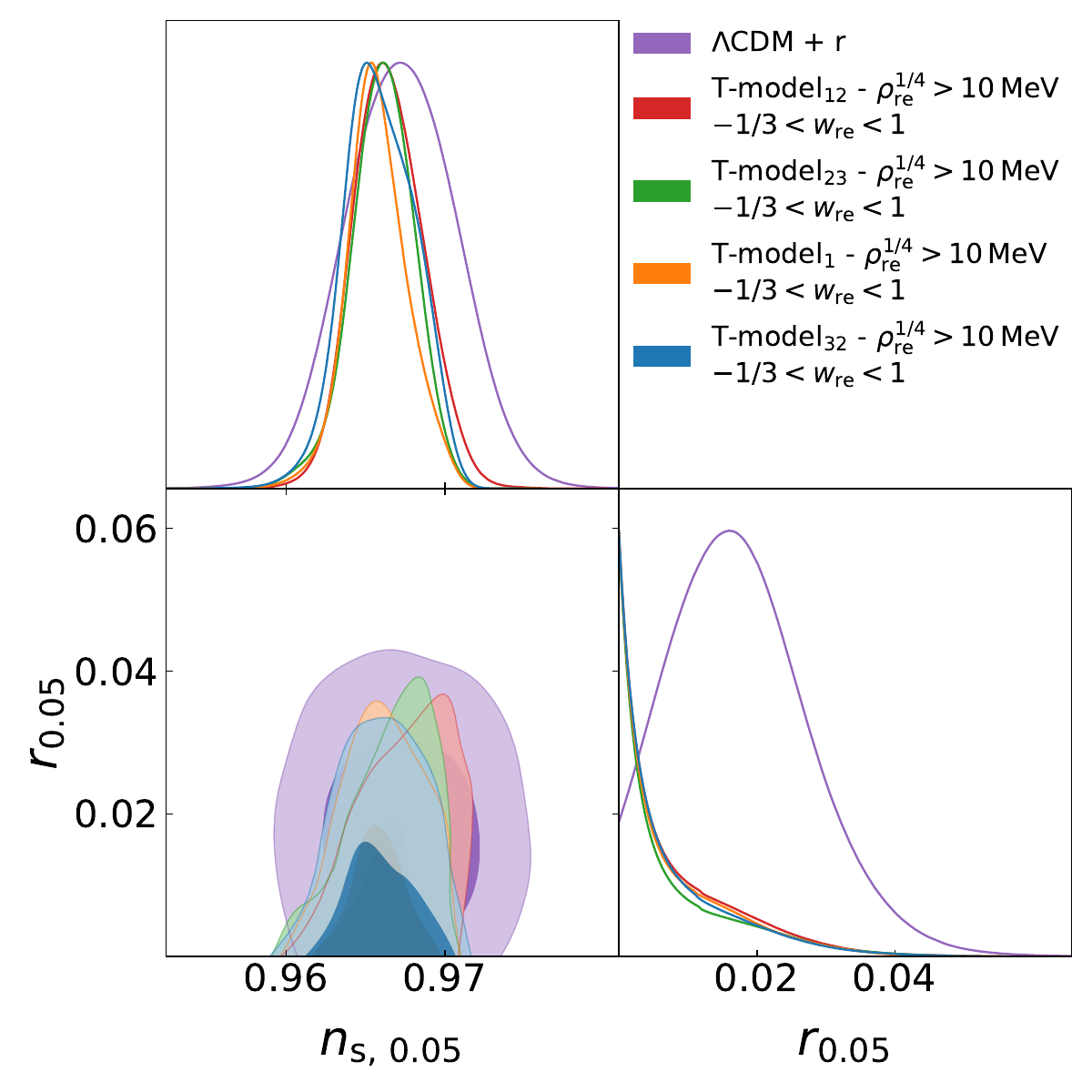} 
\\
\includegraphics[width=0.49\textwidth]{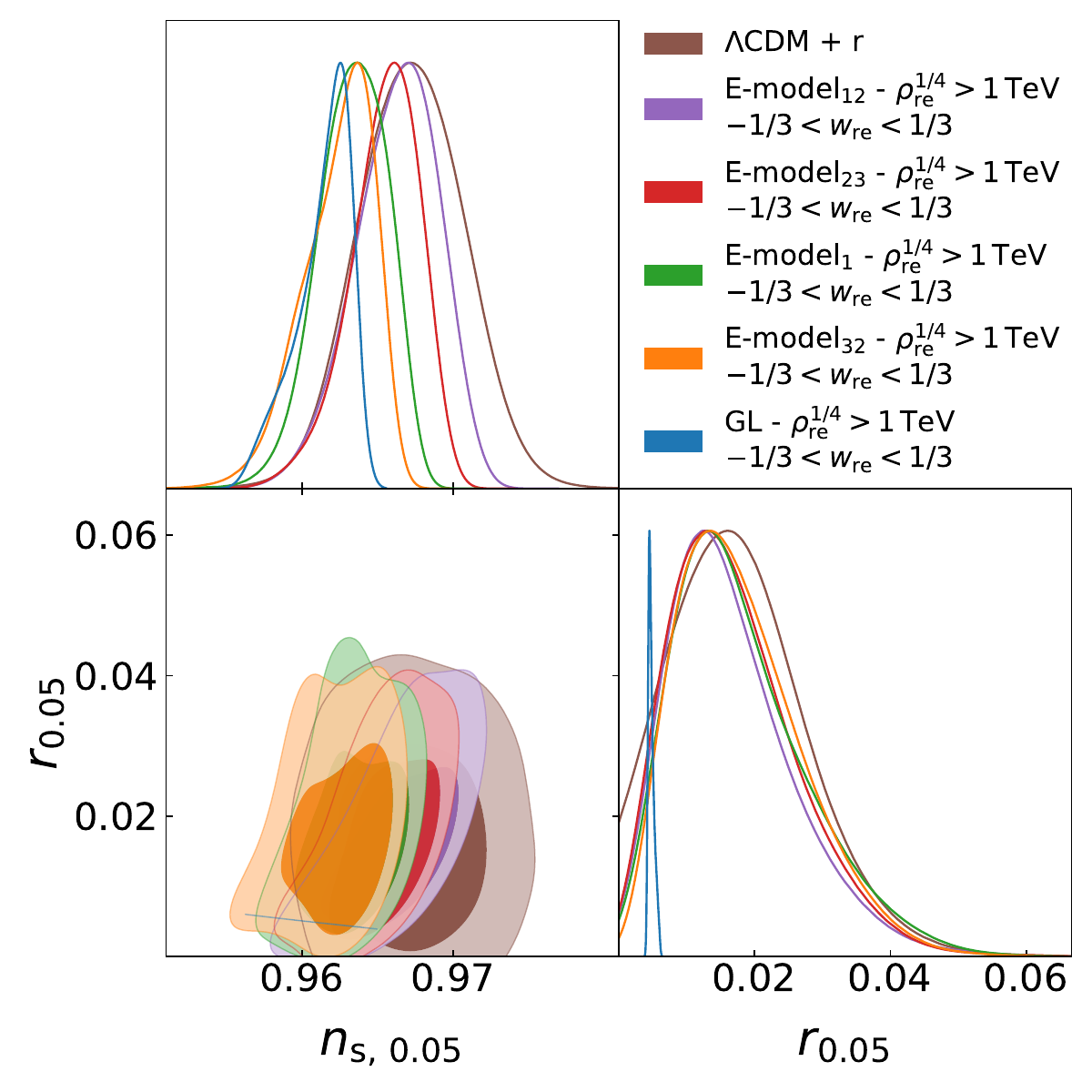}
\includegraphics[width=0.49\textwidth]{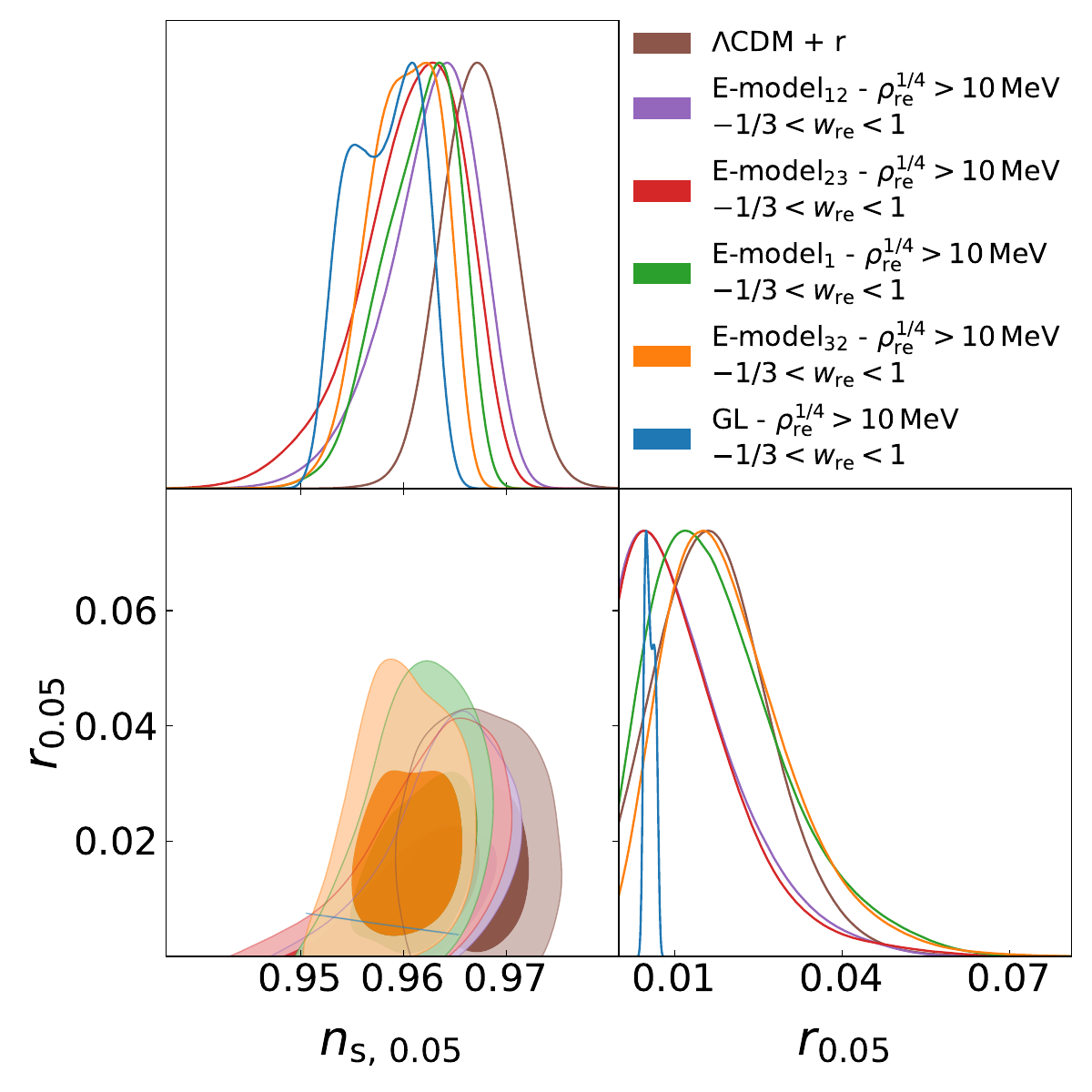}
\caption{Marginalised joint confidence contours for the scalar spectral index $n_{\rm s,\,0.05}$ and tensor-to-scalar ratio 
$r_{0.05}$ for T-model (top panels) and for E-model (bottom panels) of $\alpha$-attractor inflation at 68\% CL and 95\% CL for 
the restrictive reheating scenario on the left ($-1/3 < w_{\rm re} < 1/3,\, \rho^{1/4}_{\rm re} > 1\,{\rm TeV}$) and the 
permissive reheating one on the right ($-1/3 < w_{\rm re} < 1,\, \rho^{1/4}_{\rm re} > 10\,{\rm MeV}$).
\label{fig:nsr_wre_attractor}}
\end{figure}

\begin{figure}
\centering
\includegraphics[width=0.49\textwidth]{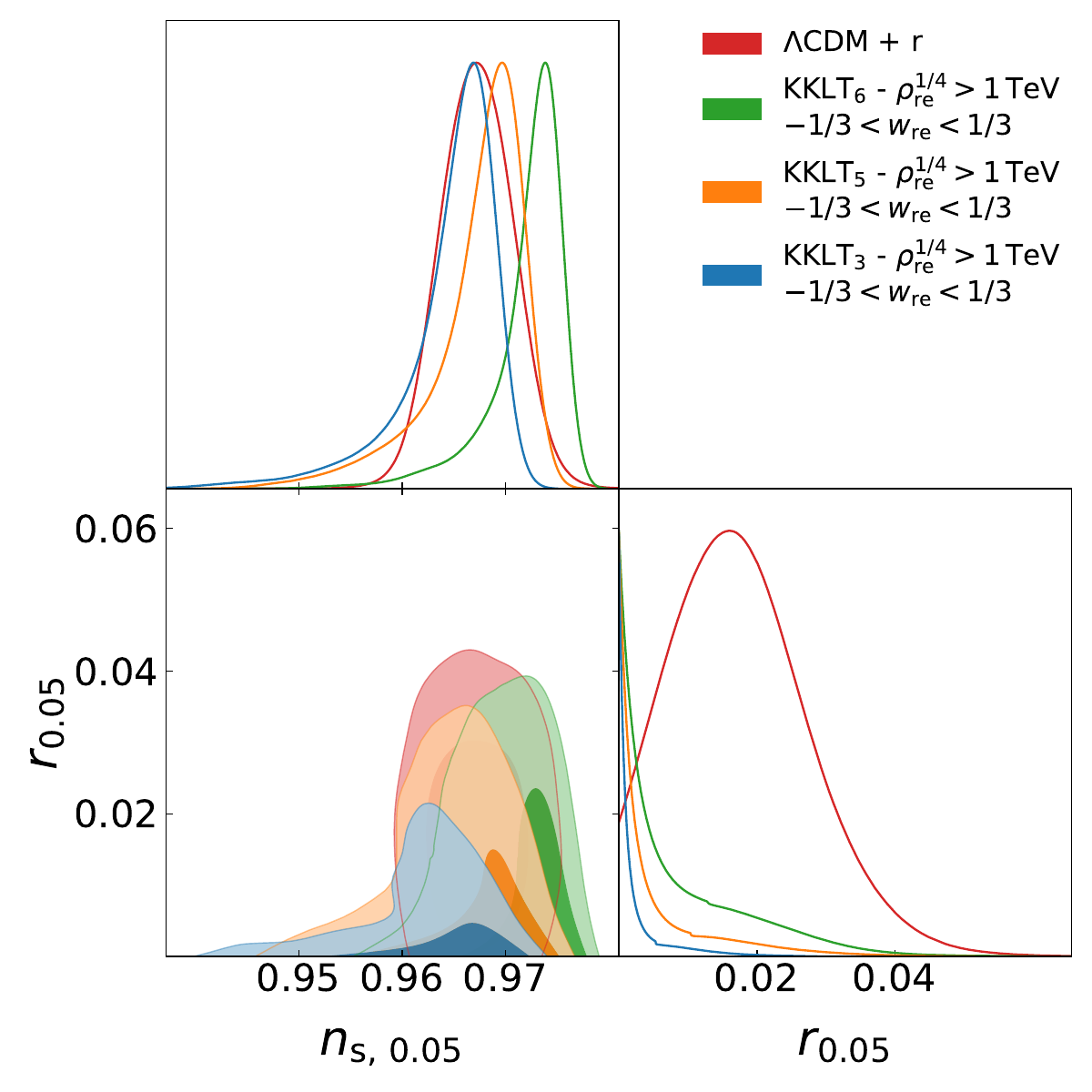}
\includegraphics[width=0.49\textwidth]{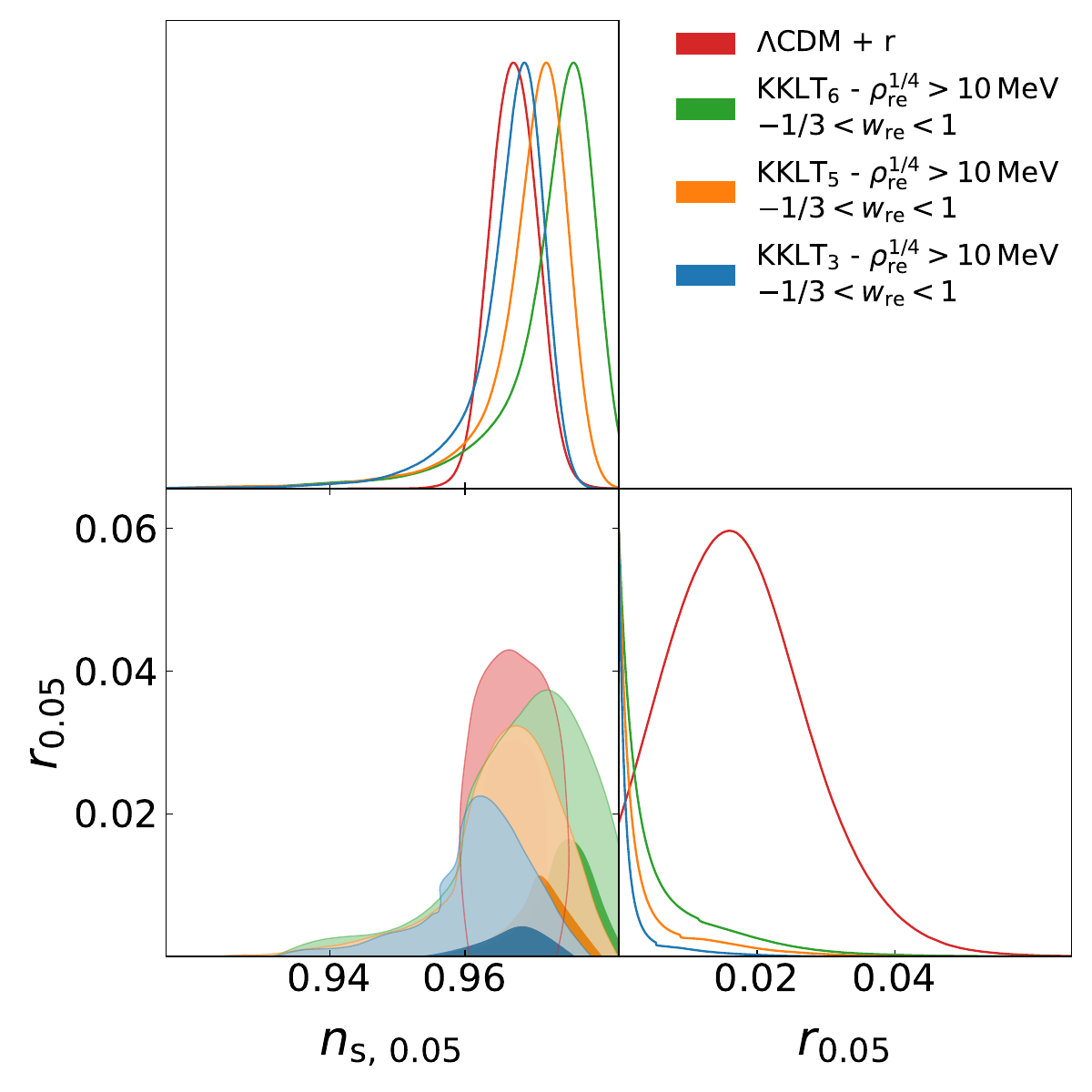}
\caption{Same as Fig.~\ref{fig:nsr_wre_attractor} for KKLT inflation.\label{fig:nsr_wre_KKLT}}
\end{figure}

We find that the results for $R+R^2$ and $\alpha$-attractor inflation with the restrictive reheating scenario are 
close to the ones obtained fixing $w_{\rm re} = 0$; see Figs.~\ref{fig:nsr_wre_R2} and \ref{fig:nsr_wre_attractor}. 
For these models a short reheating period is enough 
to fit cosmological observations. Moreover, these models would need a larger value of $e$-folds to be able to fit a larger value 
of the scalar spectral index. The situation is different for KKLT inflation where a longer reheating period 
is preferred in order to compensate for the higher value of the scalar spectral index predicted; see Fig.~\ref{fig:nsr_wre_KKLT}. 
Moving to the permissive reheating scenario all $\alpha$-attractor and KKLT models are able to cover the whole 
allowed $n_{\rm s}$-$r$ parameter space; this is mostly driven by the larger prior on $w_{\rm re}$.\footnote{Note that the range $1/3 < w_{\rm re} < 1$ is less plausible but possible, see Ref.~\cite{Pallis:2005bb}.
For these stiff values, a blue tilted relic of gravitational waves is expected to be generated and it could be constrained by future gravitational-wave observatories \cite{Mishra:2021wkm,Soman:2024zor}.}
Results for $R+R^2$ inflation do not change much since inflationary predictions are bounded to run only along  
the constrain equation $r \approx 3(1 - n_{\rm s})^2$; the model has no extra parameters to relax this relation.

Finally, we derive also the constraints on the temperature at the end of reheating using Eq.~\eqref{eqn:rhore}. 
As for the reheating energy density, this parameter result often unconstrained with current cosmological data, see Tables~\ref{tab:results_wrest} 
and \ref{tab:results_wperm}.
Note that, supersymmetric theories, such as those associated with $\alpha$-attractors, are affected by the {\em gravitino problem}. 
To avoid this problem, it is crucial to restrict the reheating temperature to higher than $10^9\,{\rm GeV}$ to prevent 
the overproduction of gravitinos \cite{Ellis:1982yb,Ellis:2021kad}. This lower bound, compatible with our findings, 
can be used to further tighter the prior ranges of the model inflationary parameters.

\section{Conclusions} \label{sec:conclusions}
The recent advancements in precision cosmology, particularly driven from the combination of {\em Planck} legacy data and new upper 
limits on B-modes from BICEP/Keck \cite{Planck:2018jri,BICEP:2021xfz,2022PhRvD.106h3528P}, represent a significant step forward in 
refining constraints on inflationary models.\footnote{See Refs.~\cite{Tristram:2021tvh,Galloni:2022mok,Galloni:2024lre} for analysis 
based on the combination of BICEP/Keck data and post-legacy reanalysis of {\em Planck} data, namely the {\em Planck} PR4 
\cite{Rosenberg:2022sdy,Tristram:2023haj}.}
Predictions for inflationary parameters, such as the scalar spectral index and the tensor-to-scalar ratio, 
are intricately linked through consistency relations, offering valuable insights into the dynamics of early Universe expansion.

Looking ahead, the next decade will likely see improvements primarily in the precision of constraints on the scalar 
spectral index and the tensor-to-scalar ratio. Expectations for future measurements suggest a potential threefold 
improvement in the error bars of the scalar spectral index $n_{\rm s}$ at best from ground-based CMB experiment, such as 
Simons Observatory and CMB-S4 \cite{SimonsObservatory:2018koc,CMB-S4:2020lpa}, and from their combination with large-scale 
structures experiments such as {\em Euclid} \cite{Ballardini:2016hpi,Euclid:2021qvm,Euclid:2024yrr}.\footnote{The combination of 
future small-scale CMB experiments and future galaxy surveys is also expected to largely improve the constraints on the running 
of the scalar spectral index $\alpha_{\rm s}$, see Refs.~\cite{Ballardini:2016hpi,Bahr-Kalus:2022prj,Easther:2021rdg}.} 
Conversely, advancements in B-mode polarisation measurements in the CMB anisotropies \cite{Kamionkowski:2015yta}, such as those 
expected from the LiteBIRD satellite \cite{LiteBIRD:2022cnt}, hold 
promise for significantly constraining the tensor-to-scalar ratio $r$, potentially by several orders of magnitude 
\cite{SimonsObservatory:2018koc,CMB-S4:2020lpa,LiteBIRD:2022cnt}.

In our study, we investigated the implications of recent BICEP/Keck measurements in combination to {\em Planck}'s ones 
for a selection of inflationary models, 
including $R+R^2$, $\alpha$-attractor, and D-brane inflation models. 
The models considered completely cover the $n_{\rm s}$-$r$ parameter space allowed by {\em Planck} and BICEP/Keck data 
all the way down to $r = 0$ \cite{Kallosh:2021mnu}, resulting also as good candidates to be targeted by future CMB experiments 
\cite{Kallosh:2019eeu,Kallosh:2019hzo}. By deriving the scalar spectral index and the tensor-to-scalar ratio up to second 
order in slow-roll and considering reheating uncertainties, we provided insights into the compatibility of these models 
with CMB observations. 

Our analysis revealed the importance of combined constraints on $n_{\rm s}$ and $r$ to disentangle different inflationary 
models as well as the importance to include the theoretical information on the reheating phase to shrink the predicted parameter 
space. Indeed, reheating uncertainties and uncertainties on inflationary parameters can be further reduced injecting in 
the analysis the information on the energy density distribution and equation of state of the universe between the end of 
inflation and the onset of radiation domination based on numerical simulations of the reheating epoch 
\cite{Lozanov:2016hid,Antusch:2020iyq,Antusch:2021aiw}.
Additional insights into the reheating epoch can be derived from the imprints on the stochastic gravitational wave background, 
as highlighted in studies exploring the dynamics of inflationary models and the generation of gravitational waves during reheating 
\cite{2021MNRAS.502L..11V,2023JHEAp..39...81V}.

Of course, one should remember that exact predictions of these models not only depend on details of the models and 
mechanism of reheating, the addition of different datasets can shift (mostly along the $n_{\rm s}$ direction) the 
position of the allowed region \cite{Giare:2023wzl}. For instance, the addition of recent DESI DR1 galaxy and quasar 
BAO to {\em Planck} data leads to a higher value of the scalar spectral index $n_{\rm s} = 0.9700 \pm 0.0036$ at 68\% CL 
\cite{DESI:2024mwx,Wang:2024hks}, eventually going in the direction of preferring D-brane inflationary models, 
while ACT DR4 data points to even larger values as $n_{\rm s} = 1.008 \pm 0.015$ at 68\% CL \cite{ACT:2020gnv}.

In conclusion, if future measurements align with the current maximum likelihood values for $r$, and if inflation proceeded 
through a single-field slow-roll mechanism, detecting non-zero values for the running of spectral indexes and tensor 
spectral indexes may pose a challenge to the prevailing inflationary paradigm. Continued advancements in B-mode 
measurements are expected to provide further insights into the inflationary parameter space.

\acknowledgments
The author is grateful to Andrei Linde and Renata Kallosh for useful comments. 
I acknowledge financial support from the INFN InDark initiative and from the COSMOS network ({\tt www.cosmosnet.it}) 
through the ASI (Italian Space Agency) Grants 2016-24-H.0, 2016-24-H.1-2018, 2020-9-HH.0 (participation in LiteBIRD phase A). 
I also acknowledge financial support by ``Bando Giovani anno 2023 per progetti di ricerca finanziati con il contributo 5x1000 anno 2021''.

\newpage
%%%%%%%%%%%%%%%%%%%%%%%%%%%%%%%%%%%%%%%%%%%%%%%%%%%%%%%%%%%%%%%%%%%%%%%%%%%%%%%
\appendix
%%%%%%%%%%%%%%%%%%%%%%%%%%%%%%%%%%%%%%%%%%%%%%%%%%%%%%%%%%%%%%%%%%%%%%%%%%%%%%%

\section{Additional tables}
We collect constraints and mean values on the sampled and derived inflationary parameters for the restrictive 
reheating scenario in Table~\ref{tab:results_wrest} and for the permissive one in Table~\ref{tab:results_wperm}.

\begin{table*}[!hb]
\centering
\begin{tabular}{l|cccc}
\hline
\hline
Parameter                                        & $\Lambda$CDM+$r$              & $R+R^2$      & GL    & Poincar\'e$_{7/3}$ \\
\hline
$\ln \left(  10^{10} A_{\rm s} \right)$          & $3.048^{+0.012}_{-0.014}$  & $3.050^{+0.012}_{-0.014}$ & $3.049^{+0.012}_{-0.013}$   & $3.049\pm 0.013$ \\
$\ln \left( \rho_{\rm re}/M_{\rm Pl}^4 \right)$ (at 95\% CL) & $-$            & $-$  & $-$   & $-$ \\
$w_{\rm re}$ (at 95\% CL)                                    & $-$            & $-$  & $-$   & $> -0.22$ \\
\hline
$n_{\rm s,\,0.05}$                                & $0.9672\pm 0.0035$        & $0.9641^{+0.0015}_{-0.0007}$ & $0.9612^{+0.0026}_{-0.0013}$   & $0.9625^{+0.0024}_{-0.0012}$ \\
$r_{0.05}$   (at 95\% CL)                                     & $< 0.036$     & $0.0037^{+0.0006}_{-0.0004}$ & $0.0048^{+0.0011}_{-0.0007}$    & $0.0081^{+0.0016}_{-0.0012}$ \\
$N_{0.05}$                                        & $-$                       & $53.0^{+2.3}_{-1.0}$    & $51.9^{+3.5}_{-1.9}$    & $51.9^{+3.3}_{-1.9}$ \\
$\log \left(T_{\rm re}/{\rm GeV}\right)$  (at 95\% CL)                       & $-$                       & $-$ & $> 4.0$   & $-$ \\
\hline
\end{tabular}
\newline
\vspace*{.2 cm}
\newline
{\small
\begin{tabular}{l|cccc}
\hline
\hline
Parameter                                        & T-model$_{1/2}$           & T-model$_{2/3}$     & T-model$_{1}$     & T-model$_{3/2}$ \\
\hline
$\ln \left(  10^{10} A_{\rm s} \right)$          & $3.050\pm 0.012$ & $3.054^{+0.013}_{-0.016}$   & $3.049\pm 0.013$     & $3.050^{+0.013}_{-0.015}$ \\
$\ln \left( \rho_{\rm re}/M_{\rm Pl}^4 \right)$ (at 95\% CL) & $-$      & $-$  & $> -123$ & $> -121$ \\
$w_{\rm re}$     (at 95\% CL)                    & $-$      & $-$  & $> -0.20$ & $> -0.17$ \\
$\alpha^{\rm T}$ (at 95\% CL)                           & $< 8.9$     & $< 9.8$ & $< 7.4$  & $< 6.5$ \\
\hline
$n_{\rm s,\,0.05}$                                & $0.9636^{+0.0019}_{-0.0016}$ & $0.9636^{+0.0016}_{-0.0011}$ & $0.9633^{+0.0014}_{-0.0007}$    & $0.9630^{+0.0015}_{-0.0006}$ \\
$r_{0.05}$    (at 95\% CL)                                    & $< 0.025$      & $< 0.028$  & $< 0.024$   & $< 0.022$ \\
$N_{0.05}$                                        & $52.7^{+2.5}_{-1.3}$         & $53.3^{+2.1}_{-1.0}$ & $53.5^{+2.0}_{-1.0}$  & $53.4^{+2.2}_{-1.1}$ \\
$\log \left(T_{\rm re}/{\rm GeV}\right)$     (at 95\% CL)   & $-$                   & $-$            & $> 4.4$        & $> 4.6$  \\
\hline
\end{tabular}}
\newline
\vspace*{.2 cm}
\newline
{\small
\begin{tabular}{l|cccc}
\hline
\hline
Parameter                                        & E-model$_{1/2}$           & E-model$_{2/3}$     & E-model$_{1}$     & E-model$_{3/2}$ \\
\hline
$\ln \left(  10^{10} A_{\rm s} \right)$          & $3.051\pm 0.014$ & $3.052\pm 0.014$   & $3.051\pm 0.014$     & $3.048\pm 0.013$ \\
$\ln \left( \rho_{\rm re}/M_{\rm Pl}^4 \right)$ (at 95\% CL)  & $-$     & $-$ & $-$  & $-$  \\
$w_{\rm re}$ (at 95\% CL)                              & $-$     & $> -0.19$ & $> -0.21$  & $> -0.23$  \\
$\alpha^{\rm E}$  (at 95\% CL)                          & $< 36.6$     & $< 25.4$  & $< 18.5$ & $< 14.6$ \\
\hline
$n_{\rm s,\,0.05}$                                & $0.9663^{+0.0031}_{-0.0024}$ & $0.9655^{+0.0027}_{-0.0019}$ & $0.9634^{+0.0026}_{-0.0020}$    & $0.9624^{+0.0029}_{-0.0017}$ \\
$r_{0.05}$                      (at 95\% CL)                  & $0.017^{+0.018}_{-0.016}$     & $0.017^{+0.018}_{-0.016}$  & $0.018^{+0.020}_{-0.017}$  & $0.018^{+0.018}_{-0.016}$ \\
$N_{0.05}$                                        & $51.1^{+3.4}_{-2.6}$         & $51.9^{+3.4}_{-2.1}$ & $51.3^{+3.4}_{-2.7}$  & $52.1^{+3.7}_{-2.1}$ \\
$\log \left(T_{\rm re}/{\rm GeV}\right)$  (at 95\% CL)      & $-$                    & $-$          & $-$        & $-$  \\
\hline
\end{tabular}}
\newline
\vspace*{.2 cm}
\newline
\begin{tabular}{l|ccc}
\hline
\hline
Parameter                                        & KKLT$_3$                & KKLT$_5$ & KKLT$_6$\\
\hline
$\ln \left(  10^{10} A_{\rm s} \right)$          & $3.059\pm 0.015$    & $3.055\pm 0.015$   & $3.055^{+0.012}_{-0.014}$ \\
$\ln \left( \rho_{\rm re}/M_{\rm Pl}^4 \right)$ (at 95\% CL)  & $-$            & $-$             & $< -49$  \\
$w_{\rm re}$ (at 95\% CL)                              & $-$     & $-$  & $-$  \\
$m\,[{\rm M_{\rm Pl}}]$  (at 95\% CL)                                     & $< 5.8$         & $< 4.9$  & $< 6.2$  \\
\hline
$n_{\rm s,\,0.05}$                                & $0.9648^{+0.0049}_{-0.0019}$ & $0.9675^{+0.0051}_{-0.0020}$ & $0.9723^{+0.0037}_{-0.0014}$ \\
$r_{0.05}$ (at 95\% CL)                                       & $< 0.012$       & $< 0.023$  & $< 0.029$  \\
$N_{0.05}$                                        & $48^{+7}_{-3}$         & $48^{+7}_{-3}$ & $49^{+6}_{-3}$ \\
$\log \left(T_{\rm re}/{\rm GeV}\right)$ (at 95\% CL)       & $-$                   & $-$           & $< 12.4$  \\
\hline
\end{tabular}
\caption{Same as Table~\ref{tab:results_w0} for the restrictive reheating scenario.\label{tab:results_wrest}}
\end{table*}

\begin{table*}
\centering
\begin{tabular}{l|cccc}
\hline
\hline
Parameter                                        & $\Lambda$CDM+$r$              & $R+R^2$      & GL    & Poincar\'e$_{7/3}$ \\
\hline
$\ln \left(  10^{10} A_{\rm s} \right)$          & $3.048^{+0.012}_{-0.014}$  & $3.051^{+0.012}_{-0.014}$  & $3.055\pm 0.015$   & $3.053\pm 0.014$ \\
$\ln \left( \rho_{\rm re}/M_{\rm Pl}^4 \right)$ (at 95\% CL)  & $-$                        & $-$             & $-$   & $-$ \\
$w_{\rm re}$                                     & $-$                        & $> 0.068$   & $> 0.044$   & $> -0.090$ \\
\hline
$n_{\rm s,\,0.05}$                                & $0.9672\pm 0.0035$        & $0.9667^{+0.0021}_{-0.0017}$ & $0.9582^{+0.0042}_{-0.0038}$   & $0.9607^{+0.0037}_{-0.0023}$ \\
$r_{0.05}$                        (at 95\% CL)                 & $< 0.036$    & $0.0032\pm 0.0008$  & $0.0056^{+0.0015}_{-0.0014}$   & $0.0088^{+0.0025}_{-0.0020}$ \\
$N_{0.05}$                                        & $-$                       & $57.5\pm 3.7$  & $48.3\pm 4.0$   & $49.7^{+4.6}_{-3.4}$ \\
$\log \left(T_{\rm re}/{\rm GeV}\right)$  (at 95\% CL)                        & $-$                       & $-$    & $-$ & $-$ \\
\hline
\end{tabular}
\newline
\vspace*{.2 cm}
\newline
{\small
\begin{tabular}{l|cccc}
\hline
\hline
Parameter                                        & T-model$_{1/2}$           & T-model$_{2/3}$     & T-model$_{1}$     & T-model$_{3/2}$ \\
\hline
$\ln \left(  10^{10} A_{\rm s} \right)$          & $3.053\pm 0.015$ & $3.050^{+0.012}_{-0.014}$   & $3.053\pm 0.014 $     & $3.051\pm 0.012$ \\
$\ln \left( \rho_{\rm re}/M_{\rm Pl}^4 \right)$ (at 95\% CL) & $-$      & $-$  & $> -123$ & $> -121$ \\
$w_{\rm re}$     (at 95\% CL)                    & $> -0.0019$      & $> 0.060$  & $> 0.041$ & $> -0.0014$ \\
$\alpha^{\rm T}$ (at 95\% CL)                           & $< 12.3$     & $< 11.1$ & $< 9.6$  & $< 9.6$ \\
\hline
$n_{\rm s,\,0.05}$                                & $0.9665^{+0.0020}_{-0.0024}$ & $0.9661\pm 0.0020$ & $0.9657^{+0.0017}_{-0.0020}$    & $0.9659\pm 0.0022$ \\
$r_{0.05}$    (at 95\% CL)                                    & $< 0.027$      & $< 0.027$  & $< 0.026$   & $< 0.026$ \\
$N_{0.05}$                                        & $57.1^{+2.7}_{-3.6}$         & $57.4^{+2.9}_{-3.2}$ & $57.6^{+2.6}_{-3.6}$  & $58.4^{+3.6}_{-4.5}$ \\
$\log \left(T_{\rm re}/{\rm GeV}\right)$     (at 95\% CL)   & $-$                   & $-$            & $-$        & $-$  \\
\hline
\end{tabular}}
\newline
\vspace*{.2 cm}
\newline
{\small
\begin{tabular}{l|cccc}
\hline
\hline
Parameter                                        & E-model$_{1/2}$           & E-model$_{2/3}$     & E-model$_{1}$     & E-model$_{3/2}$ \\
\hline
$\ln \left(  10^{10} A_{\rm s} \right)$          & $3.055^{+0.013}_{-0.014}$ & $3.058\pm 0.013$   & $3.055\pm 0.014$     & $3.053\pm 0.014$ \\
$\ln \left( \rho_{\rm re}/M_{\rm Pl}^4 \right)$ (at 95\% CL)  & $-$     & $-$ & $-$  & $-$  \\
$w_{\rm re}$ (at 95\% CL)                              & $-0.068$     & $> 0.024$ & $> -0.036$  & $> 0.037$  \\
$\alpha^{\rm E}$  (at 95\% CL)                          & $< 26.3$     & $< 21.7$  & $< 22.1$ & $< 16.1$ \\
\hline
$n_{\rm s,\,0.05}$                                & $0.9628^{+0.0052}_{-0.0034}$ & $0.9609^{+0.0059}_{-0.0037}$ & $0.9616^{+0.0044}_{-0.0029}$    & $0.9602^{+0.0040}_{-0.0030}$ \\
$r_{0.05}$                      (at 95\% CL)                  & $< 0.034$     & $< 0.034$  & $< 0.042$  & $0.020^{+0.022}_{-0.020}$ \\
$N_{0.05}$                                        & $48.5^{+4.0}_{-4.7}$         & $48.1^{+3.7}_{-4.9}$ & $49.4^{+4.8}_{-3.6}$  & $49.4\pm 3.8$ \\
$\log \left(T_{\rm re}/{\rm GeV}\right)$  (at 95\% CL)      & $-$                    & $-$          & $-$        & $-$  \\
\hline
\end{tabular}}
\newline
\vspace*{.2 cm}
\newline
\begin{tabular}{l|ccc}
\hline
\hline
Parameter                                                    & KKLT$_3$                & KKLT$_5$ & KKLT$_6$\\
\hline
$\ln \left(  10^{10} A_{\rm s} \right)$                      & $3.052\pm 0.015$    & $3.053\pm 0.014$   & $3.052\pm 0.016$ \\
$\ln \left( \rho_{\rm re}/M_{\rm Pl}^4 \right)$ (at 95\% CL) & $-$            & $-$             & $< -64.5$  \\
$w_{\rm re}$ (at 95\% CL)                                    & $-$     & $-$  & $-$  \\
$m\,[{\rm M_{\rm Pl}}]$  (at 95\% CL)                        & $< 6.3$         & $< 4.5$  & $< 4.9$  \\
\hline
$n_{\rm s,\,0.05}$                                           & $0.9679^{+0.0048}_{-0.0021}$ & $0.9696^{+0.0066}_{-0.0026}$ & $0.9730^{+0.0074}_{-0.0027}$ \\
$r_{0.05}$ (at 95\% CL)                                      & $< 0.012$       & $< 0.019$  & $< 0.025$  \\
$N_{0.05}$                                                   & $54^{+7}_{-4}$         & $52^{+10}_{-5}$ & $53^{+10}_{-6}$ \\
$\log \left(T_{\rm re}/{\rm GeV}\right)$ (at 95\% CL)        & $-$                   & $-$           & $< 10.8$  \\
\hline
\end{tabular}
\caption{Same as Table~\ref{tab:results_w0} for the permissive reheating scenario.\label{tab:results_wperm}}
\end{table*}

\newpage
%%%%%%%%%%%%%%%%%%%%%%%%%%%%%%%%%%%%%%%%%%%%%%%%%%%%%%%%%%%%%%%%%%%%%%%%%%%%%%%
\bibliographystyle{JHEP}
\bibliography{Biblio}
%%%%%%%%%%%%%%%%%%%%%%%%%%%%%%%%%%%%%%%%%%%%%%%%%%%%%%%%%%%%%%%%%%%%%%%%%%%%%%%

\end{document}